%% file: main.tex
\documentclass[11pt, a4paper]{article} 
\usepackage[force]{feynmp-auto}
\usepackage{jheppub,multirow,relsize,slashed,textpos}
\usepackage{wasysym}
\usepackage{marvosym}
\usepackage{tikz,soul}
\usepackage{braket} 
\usepackage{physics}

\input{commands.tex}

\title{$Z^\prime$s in neutrino scattering at DUNE}

\author[a]{Peter Ballett}
\author[a]{\!\!, Matheus Hostert}
\author[a]{\!\!, Silvia Pascoli}
\author[b, c, d, e]{\!\!, Yuber F. Perez-Gonzalez}
\author[f]{\!\!, Zahra Tabrizi}
\author[f]{and Renata Zukanovich Funchal}

\affiliation[a]{Institute for Particle Physics Phenomenology, Department of
Physics, Durham University, South Road, Durham DH1 3LE, United Kingdom}

\affiliation[b]{ICTP South American Institute for Fundamental Research
  \& Instituto de F\'isica Te\'orica,\\
  Universidade Estadual Paulista, Rua Dr.\ Bento T.\ Ferraz 271,
  CEP.\ 01140-070, S\~ao Paulo, Brazil}
  
\affiliation[c]{Theoretical Physics Department, Fermi National Accelerator Laboratory, P.O. Box 500, Batavia, IL 60510, USA}

\affiliation[d]{Department of Physics and Astronomy, Northwestern University, Evanston, IL 60208, USA}

\affiliation[e]{Colegio de F\'isica Fundamental e Interdisciplinaria de las Am\'ericas (COFI), 254 Norzagaray street, San Juan, Puerto Rico 00901} 

\affiliation[f]{Departamento de F\'isica Matem\'atica, Instituto de F\'isica, 
Universidade de S\~ao Paulo, C. P. 66.318, 05315-970 S\~ao Paulo, Brazil}

\emailAdd{peter.ballett@durham.ac.uk}
\emailAdd{matheus.hostert@durham.ac.uk}
\emailAdd{yfperezg@northwestern.edu}
\emailAdd{ztabrizi@if.usp.br}
\emailAdd{zukanov@if.usp.br}

\preprint{
\begin{flushright}
IPPP/18/89\\ FERMILAB-PUB-18-655-T\\ NUHEP-TH/18-12
\end{flushright}
}

\abstract{
Novel leptophilic neutral currents can be tested at upcoming neutrino oscillation experiments using two complementary processes, neutrino trident production and neutrino-electron ($\nu-e$) elastic scattering. Considering generic anomaly-free $U(1)$ extensions of the Standard Model, we discuss the characteristics of $\nu-e$ scattering as well as $e^+e^-$ and $\mu^+\mu^-$ trident production at the DUNE near detector in the presence of such BSM scenarios. We then determine the sensitivity of DUNE in constraining the well-known $L_e - L_\mu$ and $L_\mu - L_\tau$ models. We conclude that DUNE will be able to probe these leptophilic models with unprecedented sensitivity, covering unproved explanations of the $(g-2)_\mu$ discrepancy.
}

\begin{document} 

\maketitle

\section{Introduction}

The discovery of neutrino oscillation is the first laboratory-based proof of physics Beyond the Standard Model (BSM) establishing that, in contrast to the predictions of the Standard Model (SM), the neutrino sector has at least three mass eigenstates distinct from the flavour states defined by the charged-leptons. However, the mechanism which generates neutrino masses remains unknown and many competing candidate theories exist, ranging from the simplicity of a Dirac mass term protected by a symmetry (see \eg\ \cite{Chulia:2016ngi,Ma:2014qra,Aranda:2013gga}) or the popular see-saw mechanisms \cite{Minkowski:1977sc,Mohapatra:1979ia,GellMann:1980vs,Yanagida:1979as,Lazarides:1980nt,Mohapatra:1980yp,Schechter:1980gr,Cheng:1980qt,Foot:1988aq} to proposals with a more elaborate spectrum of particles. In general, more elaborate scenarios have additional motivations, including the explanation of lepton mass and flavour hierarchies (see \eg\ \cite{King:2013eh}), the matter-antimatter asymmetry of the universe \cite{Fukugita:1986hr, Asaka:2005pn, Asaka:2005an}, the existence of dark matter \cite{Boehm:2006mi,Ma:2006km}, the scale of neutrino masses \cite{Gabriel:2006ns,Davidson:2009ha,Bertuzzo:2017sbj} or anomalous experimental results \cite{Nath:2016mts}. Uncovering the nature of new physics in the neutrino sector, and its connection to other BSM concerns, will be a central aim of the experimental and theoretical programs over the next few decades.

Although significant progress has already been made, the neutrino sector remains relatively poorly explored. There are large uncertainties on the masses and mixing parameters of the light neutrinos \cite{Esteban:2018azc}, but even beyond the effects of neutrino mass, many SM cross sections are poorly known theoretically and infrequently measured. This is in part due to the typical energy scales of neutrino experiments which necessitate the mo\-del\-ling of the neutrino-nucleus interactions, but also because of the rareness of neutrino scattering (see \refref{Conrad:1997ne}). Much effort has gone in to measuring crucial cross sections at oscillation experiments \cite{Abe:2015biq,Adamson:2009ju,Aguilar-Arevalo:2013dva} and at the Main Injector Experiment for $\nu$-A (MINER$\nu$A) \cite{Ren:2017xov}, a dedicated cross section experiment. However, given the necessity and potential richness of BSM physics in the neutrino sector, and the wide array of measurements yet to be made, it is conceivable that new physics will also manifest itself as detectable signatures in neutrino scattering. It is crucial to keep an open mind about what future experimental work might find, for instance, in the auxiliary physics program of the Near Detector (ND) of the next-generation Deep Underground Neutrino Experiment (DUNE) \cite{Acciarri:2016crz}.

Novel interactions in the neutrino sector have been proposed for a variety of reasons, including as a potentially observable effect in the neutrino oscillation probabilities (see \eg\ \cite{Blennow:2016jkn}), as a way of ameliorating tension introduced by sterile neutrinos in the early universe~\cite{Hannestad:2013ana, Dasgupta:2013zpn,Mirizzi:2014ama,Cherry:2016jol,Capozzi:2017auw,Denton:2018dqq,Chu:2018gxk,Esmaili:2018qzu}, and as a possible explanation of anomalous results at short baseline~\cite{Bertuzzo:2018itn,Ballett:2018ynz,Arguelles:2018mtc}. Models which introduce new interactions between neutrinos and matter have been discussed in simplified settings~\cite{Boehm:2004uq, Cerdeno:2016sfi, Denton:2018xmq}, via Effective Field Theory \cite{Falkowski:2018dmy,Falkowski:2019xoe} and specific UV complete models~\cite{Farzan:2015doa}.
One class of models restricts the new interactions to leptons. This arises most naturally in settings with a gauged subgroup of lepton number, with most attention given to the anomaly free subgroups $L_\alpha - L_\beta$ for $\alpha,\beta \in \{e,\mu,\tau\}$ \cite{He:1991qd,He:1990pn}. Such leptophilic interactions must satisfy strong constraints from processes involving charged-leptons \cite{Bauer:2018onh}, but in the case of a gauged $L_\mu - L_\tau$ symmetry, neutrino processes have been found to be particularly competitive \cite{Altmannshofer2014}.

In this work, we study potential constraints which can be placed on a general set of leptophilic $Z^\prime$ models in the two most likely channels for BSM scattering at the near detector  
of DUNE: $\nu-e$ scattering and $\nu\ell\ell$ trident scattering. During ten years of running, a 75-t near detector subjected to the intense neutrino beam at the Long-Baseline Neutrino Facility (LBNF) will provide tens of thousands of $\nu-e$ scattering events. The cross section for this process is theoretically well understood and can therefore be a sensitive probe of BSM physics. Additionally, this process has received special interest due to its potential in reducing systematic uncertainties in the neutrino flux \cite{Park:2013dax,Bian:2017axs}, an undertaking which can be affected by new physics. Despite not being a purely leptonic process, neutrino trident production can also be measured with reasonable precision at DUNE, where hundreds of coherent and diffractive trident events are expected at the ND \cite{Ballett:2018uuc}. We study the neutral current channels with dielectron or dimuon final states, pointing out how the new physics contribution impacts the non-trivial kinematics of these processes. The main advantage in such measurements lies in flavour structure of dimuon tridents, which can be used to constrain otherwise difficult to test models, such as the $L_\mu-L_\tau$ gauge symmetry \cite{Altmannshofer2014}.

Although these processes can place stringent bounds on many classes of mediators, many scenarios are already heavily constrained through other experimental work. A recent study of several different $U(1)_X$ models using $\nu-e$ scattering was presented in Ref.~\cite{Lindner:2018kjo}, where data from past $\nu-e$ experiments CHARM-II, GEMMA and TEXONO has been used to put bounds on the couplings and masses of general $Z^\prime$s. Novel charged particles are typically constrained to be very massive, leading to little enhancement of the charged current neutrino scattering rates. In particular, charged scalars have been considered in $\nu\ell\ell$ trident scattering in \refref{Magill:2017mps}, where it is found that trident measurements can provide competitive bounds on charged scalars, albeit only in simplified theoretical settings. The requirement of doubly charged scalars or the connection to neutrino masses introduced by the typical UV completions of such models dilutes the relevance of the trident bounds. Neutral scalars are viable, but also present challenging UV completions. Novel $Z^\prime$ interactions in $\nu\ell\ell$ trident scattering with dimuon final states have been studied in \refref{Altmannshofer2014}, where it was shown to be a promising channel to probe a $L_\mu-L_\tau$ gauge symmetry. This model was revisited in \refrefs{Kaneta:2016uyt}{Araki:2017wyg}, where the effects of kinetic mixing and the possibility of a measurement by T2K was alluded to. Finally, neutrino trident scattering with atmospheric neutrinos was shown to be sensitive to this model as well as to simplified scalar models in \cite{Ge:2017poy}. It should be noted, however, that as it was shown in \refref{Ballett:2018uuc} the Equivalent Photon Approximation (EPA) discussed in several recent studies \cite{Magill:2017mps, Kaneta:2016uyt} for the calculation of the trident cross section leads to intolerably large errors in $\nu\ell\ell$ scattering channels in the SM. For this reason, we calculate this process without making this approximation.

The structure of the paper is as follows. In \refsec{sec:models}, we describe the basic properties of the leptophilic scenarios that we will consider in this work. In \refsec{sec:signatures}, we discuss the calculation of the trident and $\nu-e$ scattering cross sections in a general model of leptophilic $Z^\prime$. In \refsec{sec:sensitivities}, we show how DUNE can place bounds on a few popular leptophilic $Z^\prime$ models, discussing our assumptions for experimental configurations and backgrounds. We make our concluding remarks in \refsec{sec:conclusion}.

\section{\label{sec:models}Leptophilic $Z^\prime$ models}

Since we are interested in models where the novel neutral currents are present only in the lepton sector, let us consider explicitly a ${\rm U(1)}_{Z^\prime}$ extension of the SM whose Lagrangian is given by
\begin{equation}\label{eq:lagrangian}
\mathcal{L} \supset -g^\prime Z^\prime_\mu \left[ Q^\text{L}_\alpha\, \overline{L_L^\alpha}\gamma^\mu L_L^\alpha  + Q^\text{R}_\alpha\, \overline{\ell_R^\alpha}\gamma^\mu \ell_R^\alpha + \sum_{\rm N} Q_\text{N}\, \overline{N_R}\gamma^\mu N_R \right],
\end{equation}
where $L_\alpha$ ($\ell_\alpha$) represents the leptonic SU(2) doublet (singlet) of flavour $\alpha\in\{e,\mu,\tau\}$, and we included ${\rm N}$ right-handed neutrinos with charges $Q_{\rm N}$ under the new symmetry for completeness. Thus, we have $7+{\rm N}$ new parameters to characterize the couplings between the new boson and the lepton sector, one gauge coupling $g^\prime$ and $6+N$ charges $\{Q_\alpha^\text{L}, Q_\alpha^\text{R}, Q_\text{N}\}$. Below the scale of the Electroweak Symmetry Breaking (EWSB), the relevant interaction terms in the \lagrangian\ are given by 
\begin{align}\label{eq:lagrangian_below_EWSB}
\mathcal{L} \supset -g^\prime Z^\prime_\mu \left[ Q^\text{L}_\alpha\,\overline{\nu_\alpha}\gamma^\mu P_\text{L}\nu_\alpha  + \frac{1}{2}\,\overline{\ell_\alpha}\gamma^\mu(Q_\alpha^\text{V} - Q_\alpha^\text{A}\gamma^5) \ell_\alpha + \sum_{\rm N} Q_\text{N}\, \overline{N_R}\gamma^\mu N_R \right],
\end{align}
where $Q^\text{V}_\alpha \equiv Q^\text{L}_\alpha + Q^\text{R}_\alpha$ and $Q^\text{A}_\alpha \equiv Q^\text{L}_\alpha - Q^\text{R}_\alpha$. We note that the right-handed singlets could modify the form of the neutrino interaction in \refeq{eq:lagrangian_below_EWSB} by introducing a right-chiral current. The details of this would depend on the relationship between these chiral states and the flavour-basis neutrino $\nu_\alpha$. However, in practice our \lagrangian\ is fully general, as the polarization effects in the neutrino beam ensure that only the left-handed charge is relevant for light-neutrino scattering experiments. 

The \lagrangian\ in \refeq{eq:lagrangian} contains all of the terms necessary for this analysis. However, when it comes to assigning specific charges to the particles, a few wider model-building considerations are worthy of discussion. In the SM, any non-vectorial symmetry would forbid the Yukawas responsible for the charged-lepton mass terms post-ESWB; similarly, possible negative implications for neutrino mass generation are expected. The precise implementation of the neutrino mass mechanism is highly model dependent, but neutrino gauge charges are not compatible with many usual realizations\footnote{If neutrino masses are thought of as coming from a Weinberg operator, it is clear that the leptonic doublet must be uncharged under any unbroken U$(1)^\prime$ group.}. Furthermore, the novel gauge boson $Z^\prime$ will also require a mass generation mechanism, and indeed this could be achieved via the means of symmetry breaking. Although each of these is an important aspect of model building, their resolution can be expected to have little impact on the phenomenology of neutrino scattering, and we will not pursue them here.
Anomaly freedom of our new symmetry, however, is a more pertinent concern. It has been shown that an anomalous group can always be made anomaly free via the introduction of exotically charged sets of fermions which can be given arbitrarily large masses \cite{Batra:2005rh}. Yet these novel fermionic states necessarily introduce effects at low-scales, which in some cases can strongly affect the phenomenology of the model \cite{Dror:2017ehi}. Therefore, while it seems likely that mass generation can be addressed with the addition of new particles which do not interfere with neutrino scattering phenomenology, anomaly freedom is more pernicious. For this reason we will briefly discuss how anomaly freedom will dictate the types of leptonic symmetries that we consider in the remainder of this work.

\paragraph{Anomaly freedom.} The most general anomaly-free symmetries compatible with the SM were first deduced in the context of Grand Unification Theories (GUT) \cite{Barger:1982bj,Barr:1986hj}. More recently, an atlas of all anomaly-free $U(1)$ extensions of the SM with flavour-dependent charges has been provided by Ref.~\cite{Allanach:2018vjg}. Interestingly, the only anomaly-free subgroups of the SM with renormalisable Yukawa sector are leptophilic: the lepton-family number differences $L_\alpha -L_\beta$ $(\alpha,\beta=e,\mu,\tau)$ \cite{He:1990pn,He:1991qd}. The popular $B-L$ symmetry is in fact anomalous unless right-handed SM singlets are added with the appropriate charges. This is well motivated by the necessity of neutrino mass ge\-ne\-ra\-tion but remains a hypothesis, as not all models of neutrino mass require novel fermionic content. 
For the sake of discussion, we follow a similar logic and consider the most general anomaly free subgroups of the SM accidental leptonic symmetries allowing for an arbitrary number of right-handed fermionic singlets. These would presumably be associated with the neutrino mass generation mechanism, but we impose no specific relations in this regard due to the significant model-building freedom. The anomaly conditions for a leptophilic model with right-handed neutrinos are given below \cite{Ellis:2017nrp} \footnote{Notice that $\rm U(1)_{Z^\prime}^3$ together with gauge-gravity conditions imply that the number of right-handed states must be at least $N=3$.}
\begin{subequations}\label{eq:anom}
    \begin{align}
        {\rm SU(2)_W^2\times \rm U(1)_{Z^\prime}} & &\sum_\alpha Q_\alpha^{\rm L} &= 0,\\
        {\rm U(1)_Y^2\times \rm U(1)_{Z^\prime}}& &\sum_\alpha\left[ \frac{1}{2}Q_\alpha^{\rm L} - Q_\alpha^{\rm R}\right]&= 0,\\
        {\rm U(1)_Y\times \rm U(1)_{Z^\prime}^2} &  &\sum_\alpha \left[(Q_\alpha^{\rm L})^2 - (Q_\alpha^{\rm R})^2\right]&= 0,\label{eq:YZp2}\\
        {\rm U(1)_{Z^\prime}^3} & & \sum_\alpha  \left[2 (Q_\alpha^{\rm L})^3 - (Q_\alpha^{\rm R})^3\right] - \sum_{\rm N}Q_{\rm N}^3&= 0,\\
        \text{Gauge-Gravity} & & \sum_\alpha \left[2Q_\alpha^{\rm L} - Q_\alpha^{\rm R}\right] - \sum_{\rm N} Q_{\rm N} &= 0.
    \end{align}
\end{subequations}

In the absence of new $N_R$ particles ($Q_{\rm N}=0$) and assuming that $Q_{\alpha}^{\rm L}=Q_{\alpha}^{\rm R}$, that is considering vector couplings, we find the three well-known discrete solutions for the Eqs.\ \eqref{eq:anom}: the antisymmetric pairs $L_\alpha- L_\beta$, $\alpha,\beta = \{e,\mu,\tau\},\ \alpha\neq\beta$. As far as anomalies are concerned, all three pairs are equal, but frequently focus falls on $L_\mu-L_\tau$, which has no coupling to electrons and correspondingly weaker constraints. If we reconsider these conditions with charged right-handed neutrinos, we find a one dimensional continuous family of potential symmetries which can be consistently gauged. We can parametrise this as
\begin{align}
     \varrho (L_\alpha - L_\beta) + \vartheta (L_\beta - L_\lambda),
\end{align}
with
\begin{align}\label{eq:Ncharges}
3\varrho\,\vartheta(\vartheta - \varrho) = \sum_{\rm N}Q_{\rm N}^3. 
\end{align}
What we have shown is that linear combinations of the $(L_\alpha - L_\beta)$ choice of charges yield an anomaly free scenario provided N right-handed neutrinos respecting \refeq{eq:Ncharges} are added to the theory. We have checked that the ``anomaly-free atlas'' in~\cite{b_c_allanach_2018_1478085} contains a subset of these solutions, which are more general.

The above conclusions are based on the assumption of vectorial charge assignments. In the SM, this requirement is a consequence of the origin of mass assuming a chargeless Higgs. However, in non-minimal models this requirement could be relaxed. Even with this extra freedom, not all charge assignments are allowed: for example, a purely chiral $U(1)^\prime$ cannot satisfy \refeq{eq:YZp2} without additional matter charged under the SM gauge group. The axial-vector case, however, does have further solutions: we find that the same one-dimensional family of charges is allowed as for the vectorial gauge boson --- in this case, the charges apply to the left-handed fields and the right-handed have opposite charges. In such a model the leptonic mass generation mechanism would necessarily be more complicated than in the SM, but such a possibility is not excluded. UV completions of an axial-vector $Z^\prime$ have been presented in \cite{Ismail:2016tod,Kahn:2016vjr}, however, these generally introduce extra bounds that are expected to be stronger than neutrino scattering bounds (see \eg\ \cite{Dror:2017nsg,Dror:2017ehi}). For this reason, we only comment on the consequences of an axial-vector case in our calculations, but do not develop any particular model or constraint.

\paragraph{Kinetic mixing.} The symmetries of our SM extensions allow for kinetic mixing between the $Z^\prime$ and the SM gauge bosons \cite{Holdom:1985ag,Kamada:2015era,Ibe:2016dir}
\begin{align}
    \mathcal{L}_{\rm mix} = -\frac{\varepsilon}{2}F_{\kappa \rho}F^{\prime\,\kappa\rho},
\end{align}
where $F_{\kappa \rho}$ and $F^{\prime\,\kappa \rho}$ are the field strength tensors of the hypercharge and the $Z^\prime$ boson, respectively. The presence of such coupling introduces a very rich phenomenology and has been explored in great detail in the literature \cite{Alexander:2016aln}. In this work, we choose to focus on the less constrained possibility of vanishing tree-level kinetic mixing. In this case, kinetic mixing is still radiatively generated due to the presence of particles charged under both the SM and the new $U(1)$ group. As well as the SM particle content, additional particles present in the UV theory may also contribute to kinetic mixing, but we will neglect these contributions in this study as they are highly model dependent \footnote{The authors of Ref.~\cite{Escudero:2019gzq} have calculated the contribution to kinetic mixing in the $L_\mu-L_\tau$ model from a pair of scalars with opposite charges. These are typically subdominant, provided the mass hierarchy between the two scalars is not much larger than that of the charged leptons.}. We compute $\varepsilon$ between the $Z^\prime$ and the SM \emph{photon}, and find the one-loop result to be finite for any $\varrho(L_\alpha - L_\beta) + \vartheta (L_\beta - L_\lambda)$ gauge group, with divergences cancelling between families. In particular, for the $L_\mu-L_\tau$ model our result is in agreement with Refs.~\cite{Araki:2017wyg,Escudero:2019gzq}
\begin{equation}
    \varepsilon(q^2) = \frac{e g^\prime}{2\pi^2} \int_{0}^{1}\dd x\, x (1-x) \ln{\frac{m_\mu^2 - x(1-x)q^2}{m_\tau^2 - x(1-x)q^2}} \,\,\xrightarrow{q^2 \to 0} \,\, \frac{e g^\prime}{12 \pi^2} \ln{\frac{m_\mu^2}{m_\tau^2}}.
\end{equation}    
Note that the finiteness of the one-loop result has important consequences for the leptophilic theories we consider. As pointed out in \refref{Bauer:2018onh}, the finiteness of $\varepsilon$ implies that one is able to forbid tree-level kinetic mixing, albeit in a model dependent manner. This happens, for instance, when embedding the new leptophilic $U(1)$ group in a larger non-abelian group $G$, which is completely independent from the SM sector. This choice of one-loop generated kinetic mixing should be seen as a conservative choice; in the absence of cancellation between tree and loop-level kinetic mixing, this yields the least constrained scenario for an $L_\mu-L_\tau$ model. Additional constraints from first-family leptons are now relevant \cite{Kamada:2015era,Araki:2015mya}, especially $\nu-e$ scattering measurements, where the strength of the constraint makes up for the loop suppression in the coupling. For neutrino trident scattering, one can safely ignore loop-induced kinetic mixing contributions in the calculation since these are either smaller than the tree-level new physics contribution or yield very weak bounds compared to other processes. 

We emphasize that if accompanied by a consistent mechanism for the generation of the $Z^\prime$ mass terms and leptonic Yukawa terms, the models we consider constitute a UV complete extension of the SM. The treatment of such scenarios lies beyond the scope of this work, but we note that if their scalar sectors are light enough they can also yield rich phenomenology at low scales \cite{Heeck:2011wj}.

\section{\label{sec:signatures}Signatures of leptonic neutral currents}

When a neutrino impinges on a detector it has only two options for BSM scattering via a leptophilic mediator. In the simplest scenario, the neutrino interacts via the new mediator with the electrons of the detection medium. In this case, there is a tree-level $\nu-e$ scattering process which would be expected to show the clearest signs of new physics. For scattering off a hadron, however, the leptophilic nature of the mediator means that the first tree-level contribution will necessarily come from a diagram which also includes at least one additional SM mediator.
Any neutrino-hadron scattering process can be embellished with the new boson to create a BSM signature. In general, the final states of these processes will be either identical to the original un-embellished process (perhaps with missing energy) or it will have an extra pair of leptons in the final state. These neutrino trilepton production processes, which we will refer to as tridents for simplicity, can be subdivided into four types:
\begin{itemize}
\item $\ell \ell \ell$ trident: $\mathcal{H} + \nu_\alpha \to \mathcal{H}^\prime + \ell^-_\alpha + \ell^+_\beta + \ell^-_\beta$
\item $\nu \ell \ell$ trident: $\mathcal{H} + \nu_\alpha \to \mathcal{H} + \nu_\beta + \ell^+_\gamma + \ell^-_\delta$
\item $\nu \nu \ell$ trident: $\mathcal{H} + \nu_\alpha \to \mathcal{H}^\prime + \ell^-_\alpha + \nu_\beta + \overline{\nu}_\beta$
\item $\nu \nu \nu$ trident: $\mathcal{H} + \nu_\alpha \to \mathcal{H} + \nu_\alpha + \nu_\beta + \overline{\nu}_\beta$
\end{itemize}

We note that these processes all occur in the SM, and so the hunt for new physics will necessarily be competing against a background of genuine SM events. Moreover, for final states with missing energy in the form of neutrinos, isolating a BSM signal would necessarily rely on spectral measurements and other backgrounds have the potential to be large. In particular, the trident production of $\nu\nu\nu$ and $\nu\nu\ell$ will be seen as contributions to the NC elastic and charged-current quasi-elastic (CCQE) processes, and we expect backgrounds to be insurmountable (see \eg\ Ref.~\cite{Kelly:2019wow} for new physics contributions to CCQE processes). The $\ell\ell\ell$ channel, on the other hand, is expected to have a much more manageable SM background. Trimuon production, for instance, has been measured in the past and provides a multitude of kinematical observables in the final state~\cite{Holder:1977gp,KayisTopaksu:2004ea}. The SM rate for this channel contains radiative photon diagrams as well as hadronic contributions~\cite{Benvenuti1977,Barish1977,Smith1978}, whilst for leptophilic neutral bosons, the dominant  contributions comes from a weak process with initial and final state radiation of a $Z^\prime$, making it a less sensitive probe of light new physics. Finally, $\nu \ell \ell$ production, the most discussed trident signature in the literature, has already been observed in the dimuon channel~\cite{Geiregat:1990gz,Mishra:1991bv,Goncharov:2001qe}. This channel is by far the most important trident process for our study, as the leptonic subdiagrams contain only weak vertices in the SM.

\subsection{Neutrino trident scattering}

\begin{figure}
    \centering
    \includegraphics[width=\textwidth]{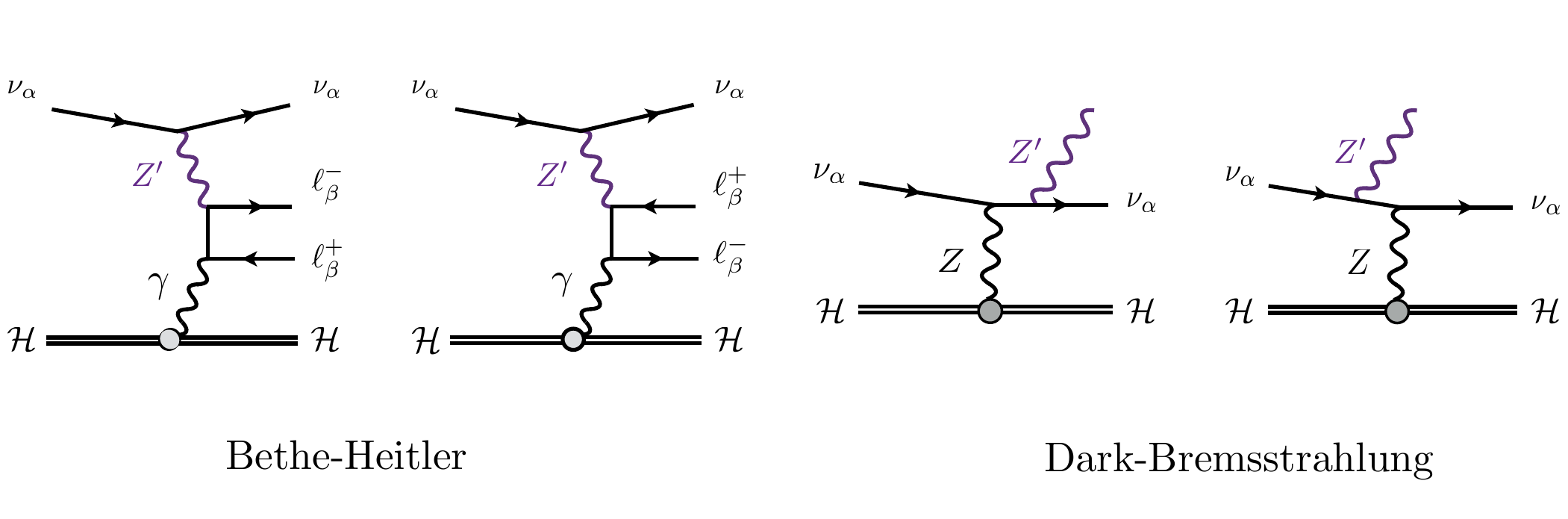}
\caption{The BSM contributions to neutrino trident production considered in our calculation. The diagrams on the top row are referred to as Bether-Heitler contributions due to their resemblance to pair-production. On the bottom row, we show diagrams with a radiative-like $Z^\prime$ contribution, which allows for the production of on-shell $Z^\prime$ particles, which subsequently decays into the charged-lepton pair. \label{fig:trident_diagrams}}
\end{figure}

In the $\nu \ell \ell$ neutrino trident scattering, an initial neutrino scatters off a hadronic target producing a pair of charged leptons in the process. Since we focus solely on neutral current processes and on flavour conserving new physics, no mixed flavour tridents are relevant and we can write 
\begin{align*} 
\mathcal{H} (P) + \nu_\alpha (p_1) \to \mathcal{H} (P^\prime) + \nu_\alpha (p_2)  + \ell^-_\beta (p_3) + \ell^+_\beta (p_4). 
\end{align*}
In the SM this process receives CC and NC contributions when $\alpha = \beta$, and is a purely NC process if $\alpha \neq \beta$. The BSM contributions to trident production we consider are shown in \reffig{fig:trident_diagrams}. Beyond computing the Bethe-Heitler (BH) contributions considered previously, we show that radiative contributions to these processes are generally small. Using the Narrow-Width-Approximation (NWA), we compute the cross section for the radiation of a $Z^\prime$ particle from a neutrino-nucleus interaction, which can then promptly decay to an $\ell^+\ell^-$ pair. We call these contributions dark-bremsstrahlung (DB) processes for their similarity with electron brehmsstrahlung in QED. We now discuss the two amplitudes individually.

\paragraph{Bethe-Heitler.} The BH amplitude can be written as follows 
\begin{equation}
\mathcal{M}_{\rm BH} = \frac{\mathrm{L}_\mu \,\mathrm{H}_{\rm EM}^{\mu}}{Q^2}.
\end{equation}
where $Q^2 \equiv -q^2 = (P - P^\prime)^2$ is the momentum transfer and $\mathrm{H}_{\rm EM}^{\mu}$ the hadronic amplitude for coherent or diffractive electromagnetic scattering
\begin{align}
{\rm H}^\mu_{\rm EM} &\equiv \langle {\cal H}(P) \vert J_\mathrm{EM}^\nu (q^2)\vert {\cal H}(P^\prime)\rangle\,.
\label{eq:Hmu1}
\end{align}
We refer the reader to Ref. \cite{Ballett:2018uuc} for the details on the treatment of the hadronic amplitude. 
The leptonic amplitude for NC scattering $\mathrm{L}_\mu$ reads
\begin{align}
\mathrm{L}_\mu & \equiv - \frac{ie G_F}{\sqrt{2}}[\bar{u}(p_2)\gamma^\tau(1-\gamma_5)u(p_1)] \times 
\bar{u}(p_4)\left[\gamma_\tau(\hat{V}_{\alpha\beta}-\hat{A}_{\alpha\beta}\gamma_5)\frac{1}{(\slashed{q}-\slashed{p}_3-m_3)}\gamma_\mu \right . \nonumber \\ 
& \left . + \gamma_\mu \frac{1}{(\slashed{p}_4-\slashed{q}-m_4)} \gamma_\tau (\hat{V}_{\alpha\beta}-\hat{A}_{\alpha\beta}\gamma_5) \right] v(p_3)\,.
\label{eq:Lmu1}
\end{align}
In writing the equation above, we have introduced effective vector and axial couplings containing SM and BSM contributions
\begin{align}
\hat{V}_{\alpha \beta} = g^{\ell_{\beta}}_{V} + \delta_{\alpha \beta} +  \frac{Q_{\alpha}^L Q_{\beta}^V}{2\sqrt{2} G_{F}} \frac{(g^\prime)^2}{K^2 + M^2_{Z^\prime}},
&\quad  
\hat{A}_{\alpha \beta} = g^{\ell_{\beta}}_{A} + \delta_{\alpha \beta} + \frac{Q_{\alpha}^L Q_{\beta}^A}{2\sqrt{2} G_{F}} \frac{(g^\prime)^2}{K^2 + M^2_{Z^\prime}}\,,
\label{eq:c_prescription_trident}
\end{align}
where $K^2 = -(p_1-p_2)^2$ and $g^{\ell_{\beta}}_{V}$'s $(g^{\ell_{\beta}}_{A}$'s) are the SM vector (axial) couplings. Note the dependence on the positive kinematic variable $K^2$ in the BSM contribution, which can lead to significant peaked behaviour in the cross section. To avoid numerical difficulties, we have modified the phase space treatment which is proposed in \cite{Czyz1964,Lovseth1971}, as shown in \refapp{app:phase_space}. 

\begin{figure}[t]
\includegraphics[width=0.49\textwidth]{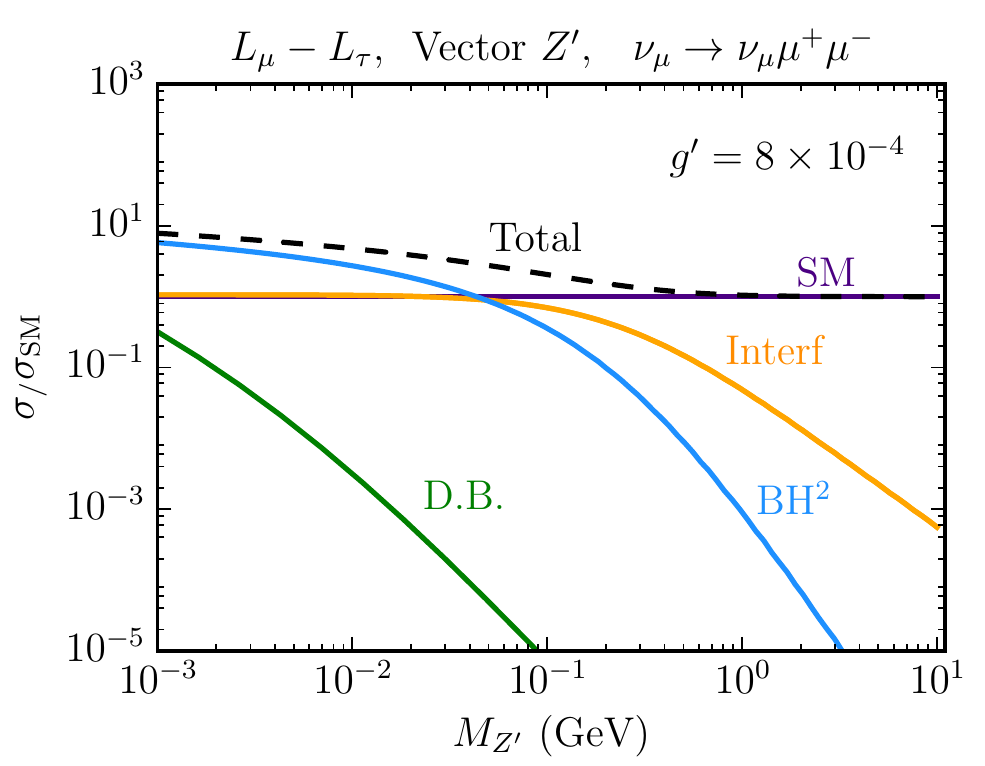}
\includegraphics[width=0.49\textwidth]{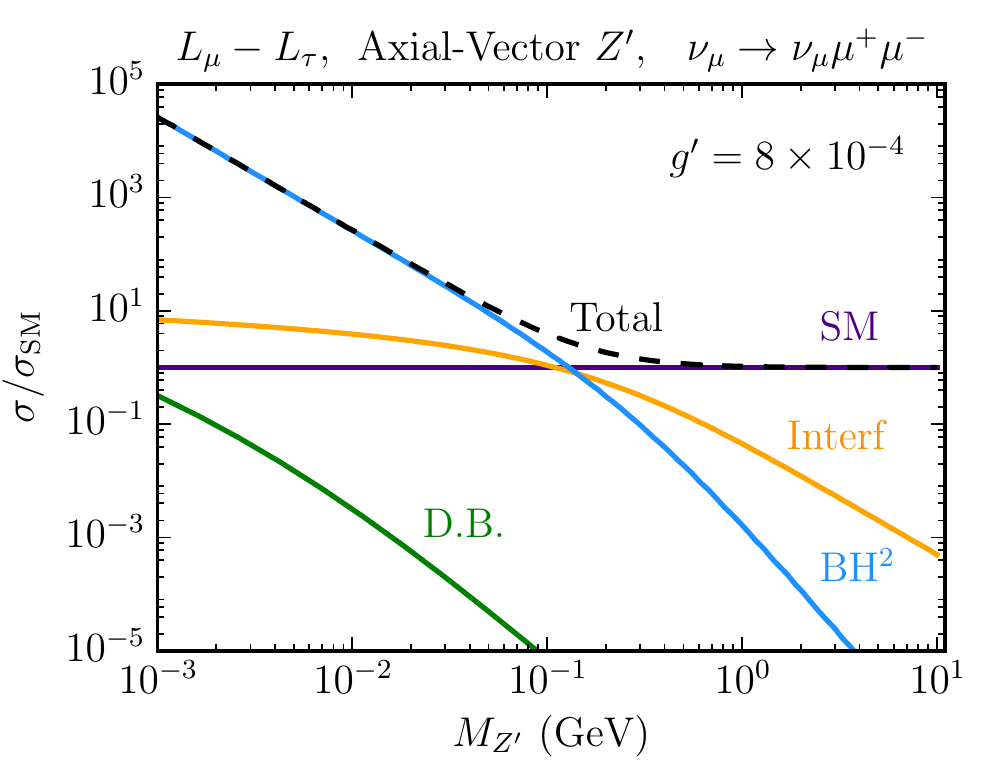}
\caption{\label{fig:xsecs_mmmm} The flux integrated cross sections normalized to the flux integrated SM trident cross section. We separate the different contributions in SM, interference between SM and Bethe-Heitler contributions (interf), Bethe-Heitler only (BH). The Dark-Bremsstrahlung (DB) cross section is also shown, but does not take the branching ratio into final state charged leptons into account.}
\end{figure}
\paragraph{Dark-Bremsstrahlung.} Due to the small decay width of the $Z^\prime$ ($\Gamma \propto g^{\prime\,2} M_{Z^\prime}$), one can obtain an estimate for its resonant production using the NWA. In the true narrow-width limit, this process reduces to a 3-body phase space calculation and does not interfere with the BH amplitude \footnote{We note that despite the fact that interference terms between resonant and non-resonant contributions vanish in the narrow-width limit, the errors induced by the NWA can no longer be shown to be of the order of $\Gamma_{Z^\prime}/M_{Z^\prime}$ \cite{Uhlemann:2008pm}. Nevertheless, we do not expect the errors of the NWA to change our conclusions.}. Our DB amplitude
for $\nu_\alpha (k_a)\,+\,  {\rm A} (k_b) \to \nu_\alpha (k_1)\,+\, Z^\prime (k_2)\,+\, {\rm A} (k_3)$ reads 
\begin{equation}
    \mathcal{M}_{\rm DB} = g^\prime Q^L_{\alpha}\, \frac{G_F}{\sqrt{2}} \, J_\mu {\rm H}_{\rm W}^{\mu},
\end{equation}%
where ${\rm H}_{\rm W}^{\mu}$ is the weak hadronic current (see \refapp{app:formfactors}) and
\begin{equation}
    J_\mu = \overline{u}(k_1) \left[ \gamma^\alpha \frac{\slashed{k_1}+\slashed{k_2}}{(k_1+k_2)^2} \gamma_\mu + \gamma_\mu \frac{\slashed{k_a}-\slashed{k_2}}{(k_a-k_2)^2} \gamma^\alpha\right] (1-\gamma^5) u(k_2) \, \epsilon^*_\alpha(k_2),
\end{equation}
where $\epsilon^*_\alpha(k_2)$ is the polarization vector of the $Z^\prime$. The previous amplitude can then be squared and integrated over phase-space for the total DB cross section. The different charged lepton final states can then be imposed with their respective branching ratio (BR). As a final remark, we note that the typical decay lengths of the new boson are typically below 1 cm for the parameter space of interest, such that their decay is indeed prompt.

\begin{figure}[t]
\includegraphics[width=0.49\textwidth]{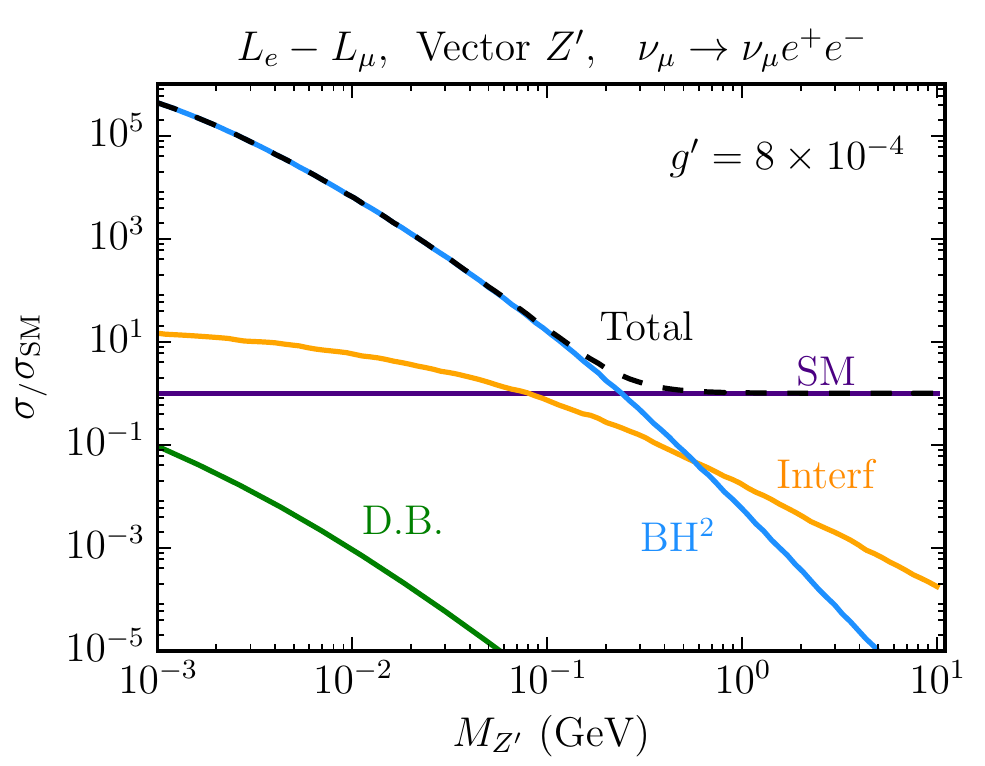}
\includegraphics[width=0.49\textwidth]{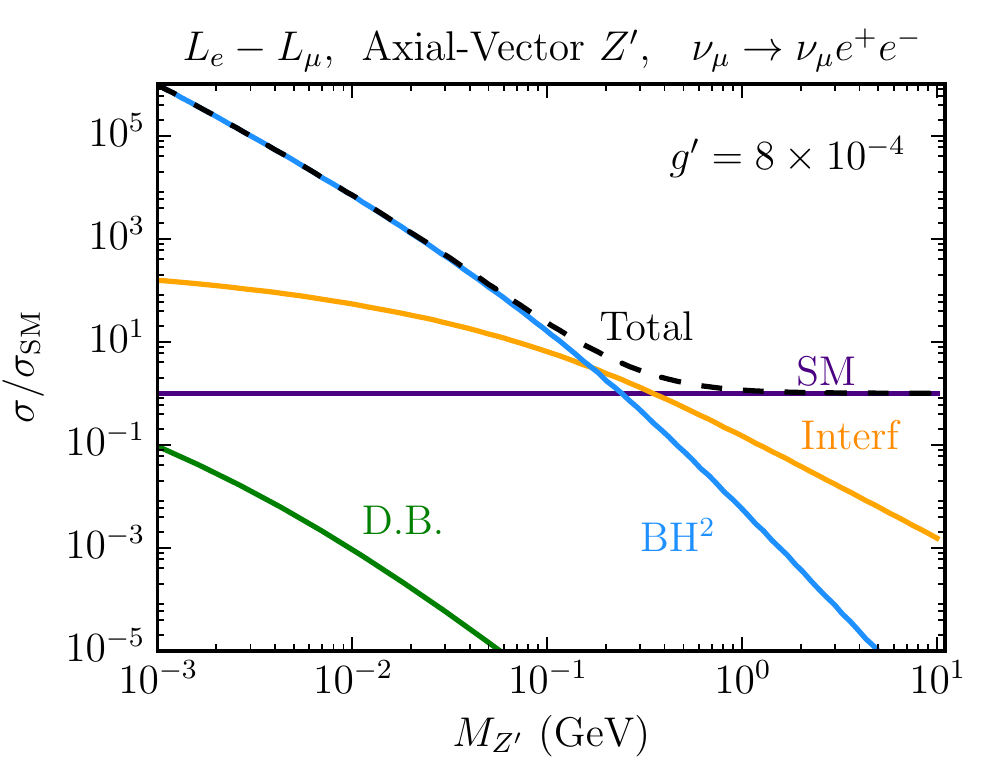}
\caption{\label{fig:xsecs_mmee} Same as \reffig{fig:xsecs_mmmm} but for the $e^+e^-$ trident channel.}
\end{figure}

From the previous discussions it is clear that the contributions with the lowest order in $g^\prime$ are the interference between the BSM and SM BH contributions, and the DB contributions. The latter, however, contains an extra power of $G_{\rm F}$ and is expected to be subdominant with respect to the interference between BH contributions. Our results for the individual flux integrated cross sections are shown in \reffig{fig:xsecs_mmmm} for $\mu^+\mu^-$ tridents and in \reffig{fig:xsecs_mmee} for $e^+e^-$ channels. We show the BH contributions as well as the DB one normalized by the SM trident cross section. All cross sections are flux integrated using the $62.4$ GeV $p^+$ DUNE flux described in \ref{sec:experimental}. For generality, we do not include the Branching Ratio (BR) factors in the DB contribution, and so the green line only applies for $\mu^+\mu^-$ tridents if $M_{Z^\prime} > 2 m_\mu$ and would suffer additional suppression due to the BR. In each figure we show two panels, one for vector couplings and one for axial-vector couplings. This is interesting from a purely computational point of view, as it shows explicitly the BH cross section scaling with the $M_{Z^\prime}$ in the two cases. Whilst the scaling is similar for dielectron tridents, it differs significantly between the vector and axial-vector cases of the dimuon cross section. This suggests the presence of mass suppression effects in the BH process. We do not investigate this further, but note that there are large cancellations between the top two diagrams in \reffig{fig:trident_diagrams} which are only present for vector-like couplings.

%
\subsubsection{The Equivalent Photon Approximation \label{sec:EPA}}

We now comment on the Equivalent Photon Approximation (EPA) for neutrino trident production. This approximation is known to perform quite badly for the SM neutrino trident production cross section \cite{Ballett:2018uuc}. One may wonder, however, if the EPA gets better or worse when computing our BSM cross sections. Naturally, it would be most inadequate for the resonant-like cross sections, since the photon propagator and the strong $1/Q^4$ behaviour is absent. However, if one focuses on the 
BH contributions, a marginal improvement of the accuracy of the approximation is seen as one lowers the mass of the $Z^\prime$ mediator. In the SM, the $\nu-\gamma$ cross sections scale as a typical weak cross section, $\sigma_{\nu \gamma} \propto G_{\rm F}^2 \hat{s}$, where $\hat{s}$ is the square of the center of mass energy of the $\nu-\gamma$ system.  On the other hand, if the cross section is dominated by the BSM BH 
contributions, 
then as we take the limit of small $Z^\prime$ masses, it scales more similarly to a QED cross section, $\sigma_{\nu \gamma} \propto 1/\hat{s}$. This behaviour, however, is only present at low masses and only for the BSM contribution. Since we are interested in regions of the parameter space where BSM and SM cross sections are of similar size, then we expect the total cross section to have a behaviour which is a combination of the two. As a sanity check, we numerically verified that for parameter space points where the BSM contributions are of the same order as the SM cross section, the improvement in the accuracy of the EPA is still not satisfactory. For instance, the ratio between the EPA prediction and the full calculation for the dimuon channel assuming a $Q_{\rm max} = (140$ MeV$)/A^{1/3}$ goes from $\approx 30\%$ in the SM to $\approx 60\%$ for $g^\prime = 8\times10^{-4}$ and $M_{Z^\prime} = 5$ MeV. For this reason, we only use the full $2 \to 4$ calculation in what follows.

\subsubsection{Trident kinematical distributions \label{sec:trident_kinematics}}
\begin{figure}[t]
\centering
\includegraphics[width=0.49\textwidth]{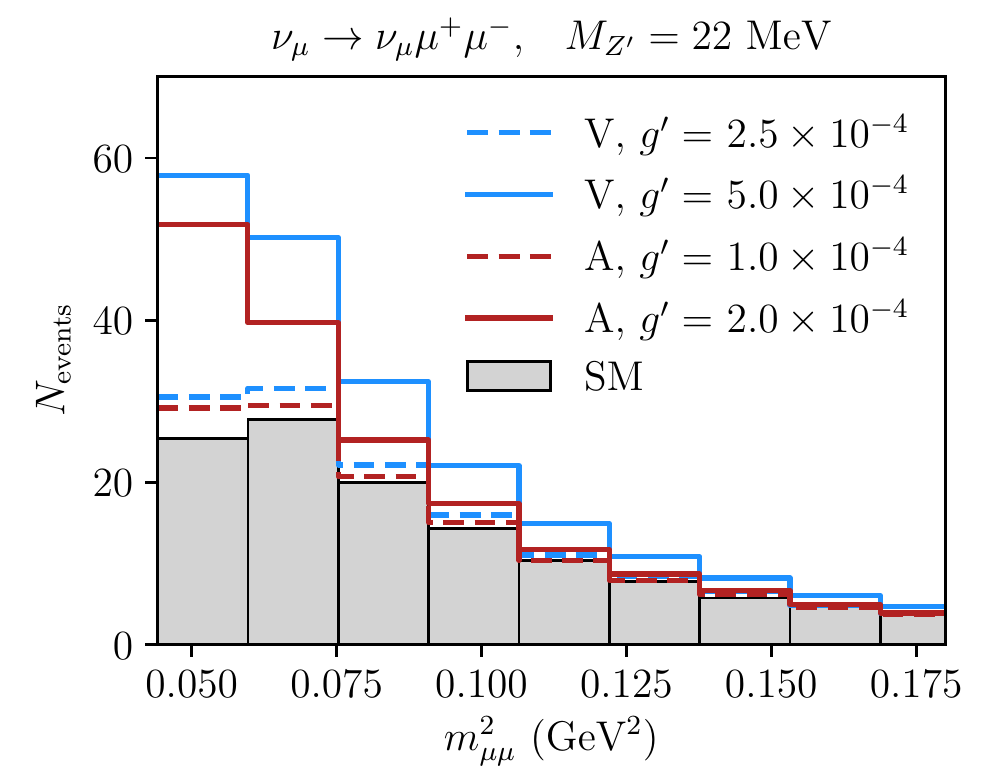}
\includegraphics[width=0.49\textwidth]{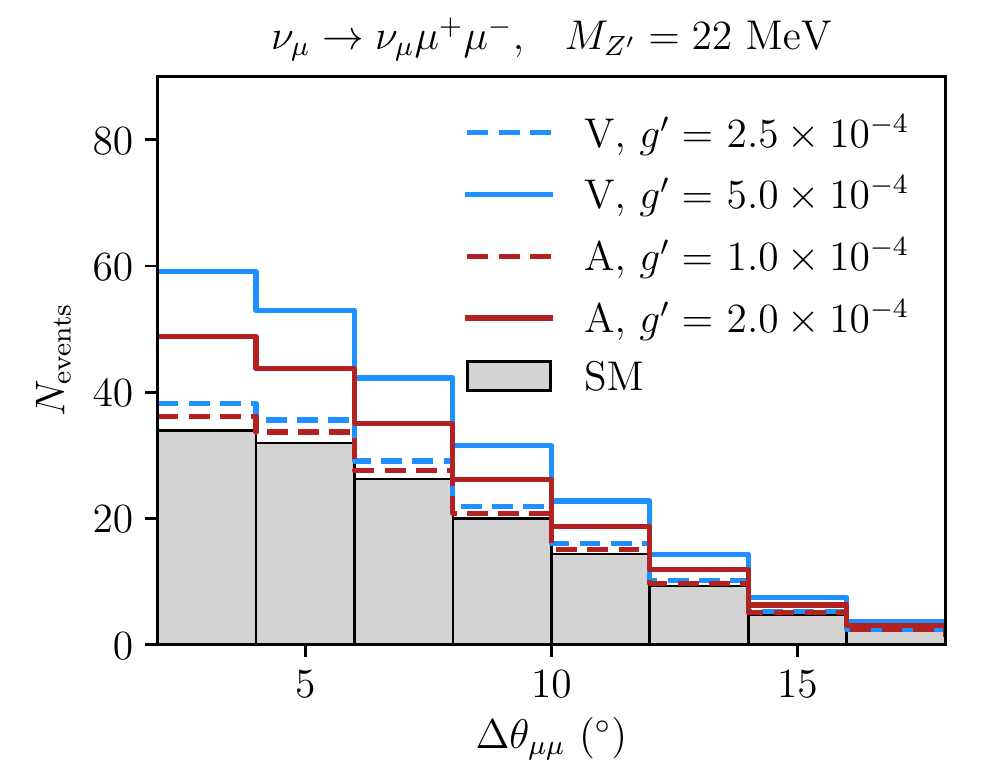}
\caption{The invariant mass (left) and charged lepton separation angle (right) flux integrated distributions for $\mu^+\mu^-$ tridents at the DUNE ND. We plot the vector (V) and axial-vector (A) cases, and assume the family charges to be just like in the $L_\mu - L_\tau$ scenario. \label{fig:mm_spectra}}
\end{figure}
The impact of new physics on the total cross section for trident production has been explored in the previous section. It is then natural to ask what the impact of new physics is on the kinematics of trident production which are, especially in the case of the invariant mass and angular variables, of utmost importance for background reduction. In this section we show how the new physics can alter the distributions in these important variables. All results that follow have been obtained using trident events produced by our dedicated Monte Carlo (MC). Smearing and selection cuts have been applied as detailed in Sec.~\ref{sec:sensitivities}.

\begin{figure}[t]
\centering
\includegraphics[width=0.49\textwidth]{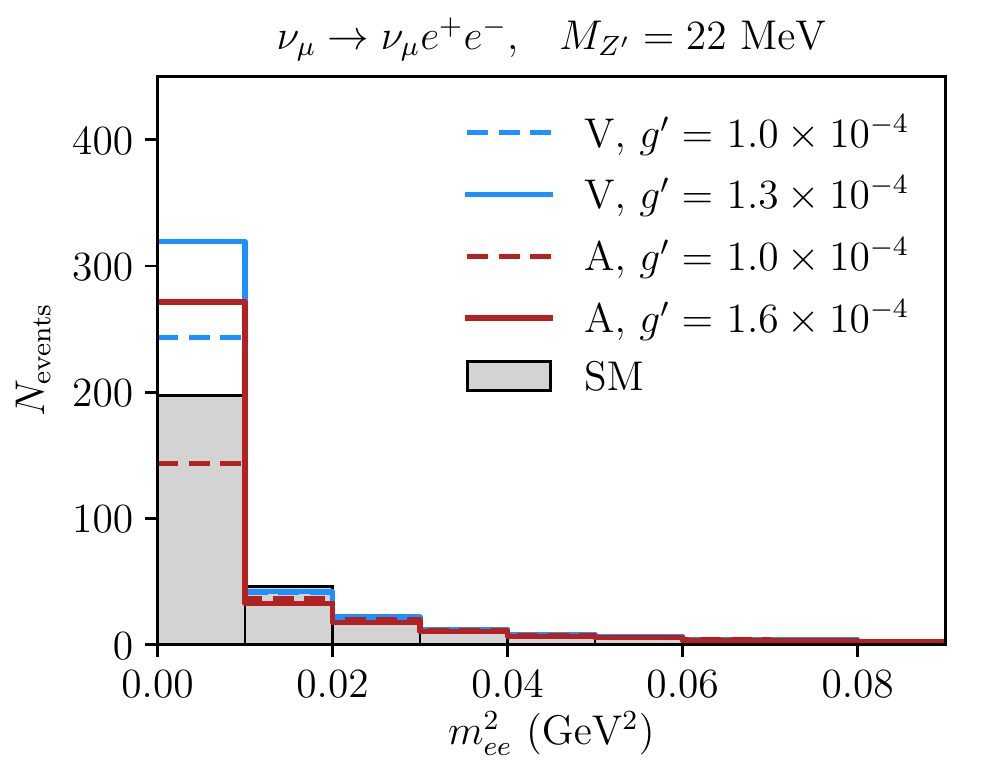}
\includegraphics[width=0.49\textwidth]{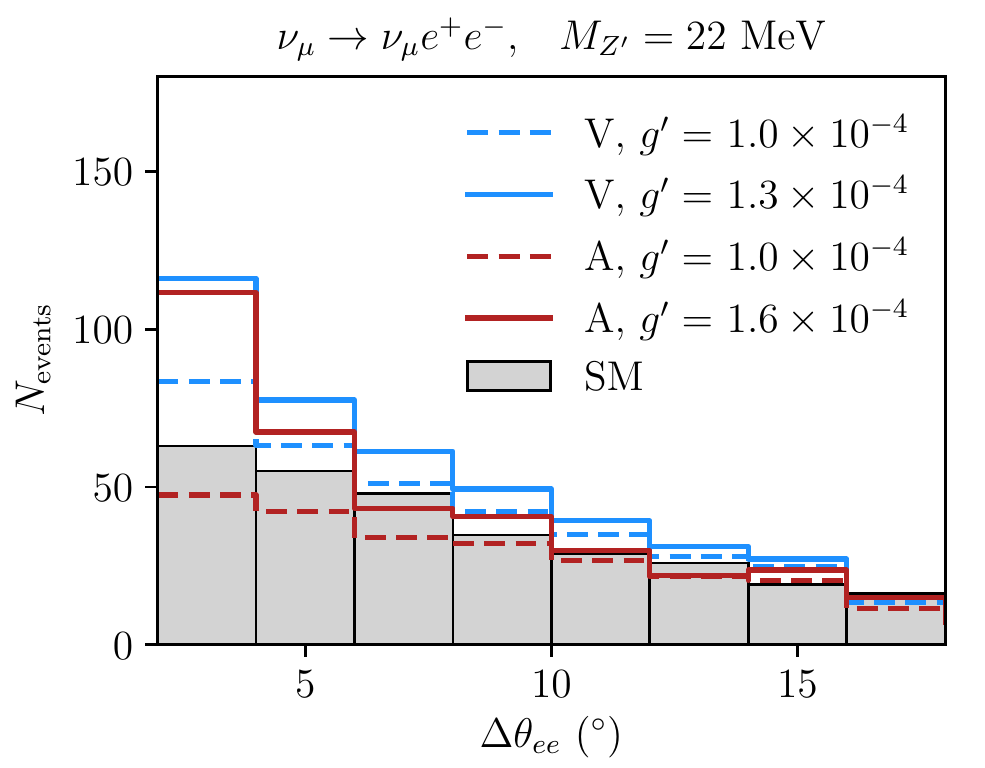}
\caption{Same as \reffig{fig:mm_spectra} but for $e^+e^-$ tridents. In all cases we assume the family charge assignment to be equal to that in an $L_e - L_\mu$ model.  \label{fig:ee_spectra}}
\end{figure}

The variables of interest in background reduction are the charged lepton invariant mass $m^2_{\ell \ell}$ and their separation angle $\Delta \theta_{\ell\ell}$. In \reffig{fig:mm_spectra} we show the dimuon invariant mass spectrum between $4 m_{\mu}^2$ and $0.2$ GeV$^2$, and the dimuon separation angle between $2^\circ$ and $18^\circ$ for a light vector boson with $M_{Z^\prime} = 22$ MeV. We show the results for the dielectron channel in \reffig{fig:ee_spectra}. The light new physics here enhances these distributions at low values of these parameters. We show our results for two types of mediators, vector and axial-vector leptophilic bosons. Comparing the couplings necessary for similar BSM enhances in the number of events, we see that axial-vector bosons lead to larger enhancements with smaller couplings. In particular, it leads to greater spectral distortions at the $Z^\prime$ masses shown.

\subsection{Neutrino-electron scattering}

Neutrino-electron scattering has long been a valuable probe of both the SM and potential new physics \cite{Marciano:2003eq,deGouvea:2006hfo,Lindner:2018kjo}. It is important to note that in the presence of novel leptophilic currents, experiments searching for $e^+e^-$ tridents would also observe anomalous $\nu-e$ event rates. In fact, given the larger statistics present in the $\nu-e$ scattering sample, this channel is expected to provide the leading constraints in our scenarios with tree-level couplings to electrons.

In order to compute the $\nu-e$ cross section in the presence of the new leptophilic interactions we need to consider an analogous modification in the NC scattering amplitude
\begin{equation}
    \mathcal{M}_{\rm \nu_\alpha-e} = - \frac{G_F}{\sqrt{2}}  \left[\bar{u}(k_2)\gamma^\mu(1-\gamma_5)u(k_1)\right] \left[\bar{u}(p_2)\gamma_\mu(C^{\rm V}_\alpha-C^{\rm A}_\alpha\gamma_5)u(p_1)\right],
\end{equation}
where the vector ($C_V$) and axial ($C_A$) effective couplings include both the SM and BSM contributions
\begin{subequations}
\begin{align}
C^{\rm V}_\alpha &= -\frac{1}{2}+2s^2_{\rm W} + \delta_{\alpha e} + \frac{Q^{\rm V}_{e} Q^{\rm L}_{\alpha} }{2\sqrt{2}G_F} \frac{(g^\prime)^2}{M_{Z^\prime}^2+2m_eT_e},\\
C^{\rm A}_\alpha &= -\frac{1}{2} + \delta_{\alpha e} + \frac{Q^{\rm A}_{e} Q^{\rm L}_{\alpha} }{2\sqrt{2} G_F} \frac{(g^\prime)^2}{ M_{Z^\prime}^2 + 2m_e T_e},
\label{eq:c_prescription_nue}
\end{align}
\end{subequations}
with, as usual, $s_{\rm W}\equiv\sin\theta_{\rm W}$, being $\theta_{\rm W}$ the weak angle and $T_e$ is the kinetic energy of the recoil electron. The loop-induced kinetic mixing in the $L_\mu-L_\tau$ model also induces a $\nu - e$ coupling
\begin{equation}
C^{\rm V}_\alpha = -\frac{1}{2}+2s^2_{\rm W} + \delta_{\alpha e} + \frac{1 }{\sqrt{2}G_F} \frac{g^\prime \, e\, \varepsilon(q^2)}{M_{Z^\prime}^2+2m_eT_e}.
\end{equation}
The differential cross section is then given by 
\begin{eqnarray}\label{nuexsection}
  \frac{d\sigma_{\nu_{\alpha}-e}}{dT_e} = \frac{2m_eG_F^2}{\pi} \left[ \left(C^{\rm L}_\alpha\right)^2 +\left(C^{\rm R}_\alpha\right)^2\left(1-\frac{T_e}{E_\nu}\right)^2-C^{\rm L}_\alpha C^{\rm R}_\alpha \, m_e \frac{T_e}{E_\nu^2} \right].
\end{eqnarray}
where the left and right handed constants are given by
\begin{align*}
C^{\rm L}_\alpha \equiv \frac{1}{2}\left(C^{\rm V}_\alpha + C^{\rm A}_\alpha\right) \qquad\text{and}\qquad
C^{\rm R}_\alpha \equiv \frac{1}{2}\left(C^{\rm V}_\alpha - C^{\rm A}_\alpha\right).
\end{align*}
For antineutrino scattering one obtains the cross section by exchanging $C_{\alpha}^{\rm L} \leftrightarrow C_{\alpha}^{\rm R}$. 

The kinetic energy of the outgoing electron is bounded by the kinematics and by the energy resolution of the detector, which effectively sets a threshold energy $T_{\rm th}$
\begin{align}
  T_{\rm th}  \le T_e \le T_{\rm max},
\end{align}
%
with $T_{\rm max} = \frac{2E_\nu^2}{m_e+2E_\nu}$, the maximum kinetic energy. We define the effective total cross section for an initial neutrino energy $E_\nu$ as
\begin{equation}\label{eq:effective_nue_xsec}
\sigma_\text{eff}(E_\nu,T_\text{th}) = \int_{T_\text{th}}^{T_{\rm max}} \frac{d\sigma}{dT_e}\,dT_e.
\end{equation}
This definition also ensures that the enhancement due to very light mediators becomes constant at around $\sqrt{2 m_e T_\text{th}}$, as discussed in \refref{Lindner:2018kjo}. This is a consequence of the detector thresholds and of the 2-body kinematics of the process. Finally, electroweak radiative corrections have been computed in the SM \cite{Bahcall:1995mm,Passera:2000ug}, but do not change our results.

\subsubsection{$\nu-e$ kinematical distributions \label{sec:nue_kinematics}}

\begin{figure}[t]
\centering
\includegraphics[width=0.49\textwidth]{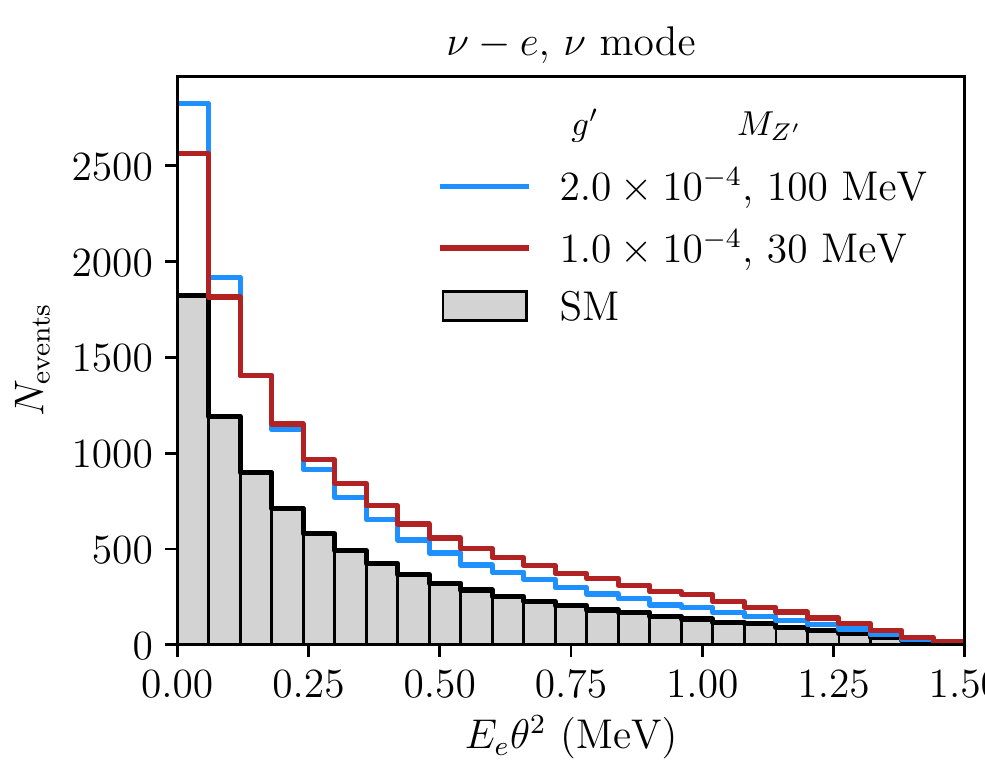}
\includegraphics[width=0.49\textwidth]{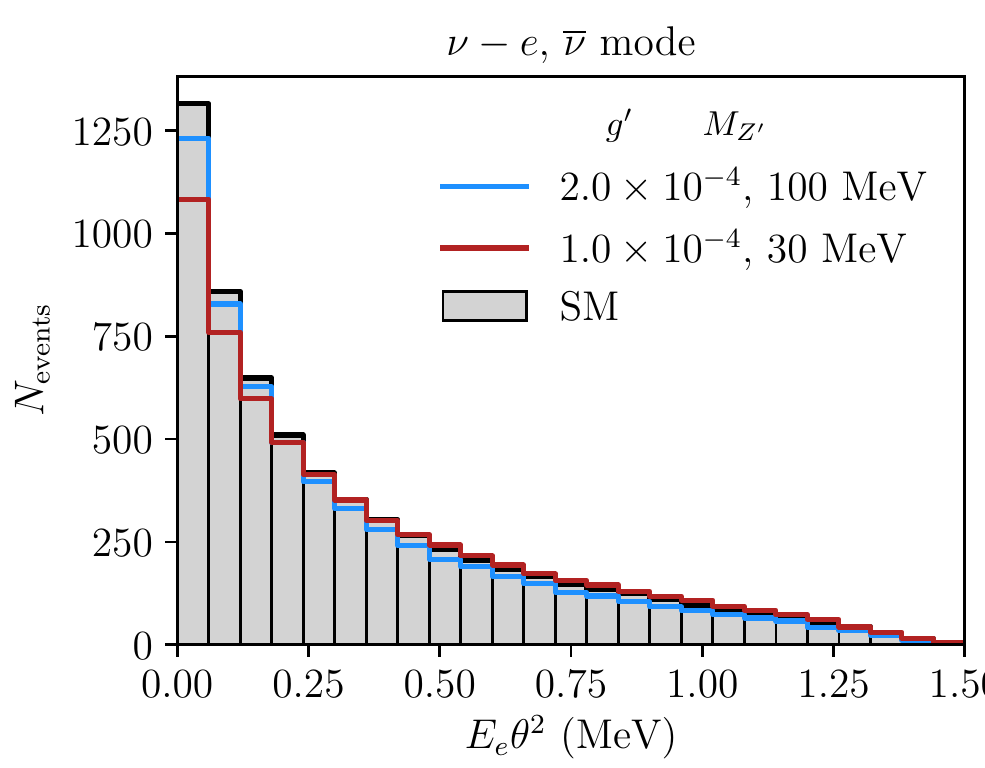}
\caption{The number of $\nu-e$ scattering events in the DUNE near detector as a function of $E_e\theta^2$ for neutrino (left) and antineutrino (right) modes. We show the SM prediction as well as two example points of new physics scenarios for a vector $Z^\prime$ with $L_e-L_\mu$ charges. Electron thresholds are taken to be $600$ MeV. \label{fig:nuedist}}
\end{figure}

The angle between the scattered electron and the outgoing neutrino $\theta$ is related to their energies as 
\begin{align*} 
1 - \cos{\theta} = m_e \frac{1 - y}{E_e}, 
\end{align*}
where $y \equiv T_e/E_{\nu}$ is the inelasticity ($T_{\rm th}/E_{\nu}<y<1$) and $E_e=T_e+m_e$ is the outgoing electron energy. This implies that at ${\cal O}(\rm GeV)$ neutrino energies, the electron recoils are very forward and obey $E_e \theta^2 < 2 m_e$, up to detector resolution. For this reason, we choose to analyse our results in terms of  $E_e \theta^2$. In this case, the differential cross section becomes
\begin{eqnarray}
  \frac{d\sigma_{\nu_{\alpha}-e}}{d(E_e\theta^2)} =\frac{E_\nu}{2m_e}\frac{d\sigma_{\nu_{\alpha}-e}}{dT_e} \Bigg|_{T_e=E_\nu(1-\frac{E_e\theta^2}{2m_e})}\,.
\end{eqnarray}
This distribution is particularly important for suppressing the background. Given the kinematics explained above, $E_e\theta^2$ must be smaller than $2m_e$ for $\nu-e$ scattering, while it is often much larger for neutrino-nucleon scattering, the dominant background (See \refsec{sec:experimental}). We show in Fig.~\ref{fig:nuedist} the expected $\nu-e$ events with respect to $E_e\theta^2$ for the SM case and a few BSM cases, in the neutrino and anti-neutrino modes of DUNE. We have used experimental resolution as described in \refsec{sec:experimental}.

\subsection{Interference effects}

Since for $\nu-e$ scattering and neutrino trident production there exists a SM contribution, we expect the experimental sensitivity to new physics to be dominated by the interference between SM and BSM contributions. We now argue what kind of interference one can expect in each one of these processes.

For neutrino trident production we follow \refref{Brown1973} and separate the differential cross section as 
\begin{equation}
d\sigma = \hat{V}^2\,\, d\sigma_{\rm V} +\hat{V} \hat{A} \,d\sigma_{\rm V-A} + \hat{A}^2 \,d\sigma_{\rm A},
\end{equation}
where we dropped the flavour indices in $\hat{V}$ and $\hat{A}$ from \ref{eq:c_prescription_trident} for simplicity. This allows us to write the interference between the SM and the vector new physics as
\begin{equation}\label{eq:trident_interference}
d\sigma_{\rm{INT}} =  \frac{Q_{\alpha}^L Q_{\beta}^V}{2\sqrt{2} G_{F}} \frac{(g^\prime)^2}{K^2 + M^2_{Z^\prime}} \left( 2 C_{\rm V}^{\rm SM}d\sigma_{\rm V}  + C_{\rm A}^{\rm SM} d\sigma_{\rm V-A}\right).
\end{equation}
Depending on the region of phase space considered, the term proportional to $d\sigma_{\rm V-A}$ can be of similar size to $d\sigma_{\rm V}$. However, $d\sigma_{\rm V-A}$ changes sign as a function of the angular variables or energies, leading to small integrated cross sections (typically two orders of magnitude smaller than the integral of the $d\sigma_{\rm V}$ term). Ignoring this term, one can then completely predict the type of interference in trident production. For $\nu_\mu \to \nu_\mu \mu^+ \mu^-$ trident production, for instance, $C_{\rm V}^{\rm SM} > 0$ and the second generation charge appears squared, leading to constructive interference in all cases. For $\nu_\mu \to \nu_\mu e^+e^-$ tridents, on the other hand, $C_{\rm V}^{\rm SM} < 0$. If the first and second generation charges come in with opposite signs, then the interference is still constructive, otherwise destructive interference happens. The same considerations also apply to antineutrino scattering if one ignores the $d\sigma_{\rm V-A}$ term. Finally, the axial-vector case is completely analogous taking $V \leftrightarrow A$ in \refeq{eq:trident_interference}.

For $\nu-e$ scattering analytical expressions can easily be used~\cite{Lindner:2018kjo}. Taking $C_L^{\rm{SM}}=-1/2+s_{\rm W}^2\sim-1/4$ and $C_L^{\rm{SM}}=s_{\rm W}^2\sim1/4$ we have
\begin{subequations}
    \begin{align}
        \frac{{d\sigma_{\rm{INT}}}_{\nu_\mu- e}}{dT_e}&\sim - \frac{\sqrt{2} m_e G_F}{4\pi}\frac{g'^2}{m_{Z'}^2+2m_eT_e}\left(-1+(1-\frac{T_e}{E_\nu})^2\right)\\
        \frac{{d\sigma_{\rm{INT}}}_{\bar\nu_\mu- e}}{dT_e}&\sim - \frac{\sqrt{2} m_e G_F}{4\pi}\frac{g'^2}{m_{Z'}^2+2m_eT_e}\left(1-(1-\frac{T_e}{E_\nu})^2\right).
    \end{align}
\end{subequations}
Since $T_e<E_\nu$, the interference term for $\nu_\mu- e$ is always positive (constructive), and for $\bar\nu_\mu -e$ it is always negative (destructive).

\section{\label{sec:sensitivities}DUNE sensitivities}

Having studied the behaviour of neutrino trident production and neutrino-electron cross sections in the presence of light new bosons, we now apply our results in sensitivity studies for the DUNE near detector. As discussed in Section \ref{sec:models}, we limit our studies to $L_e- L_\mu$ and $L_\mu- L_\tau$ models with vector gauge bosons. We start with a discussion on the experimental details, highlighting the challenges of backgrounds and laying out our statistical methods in \refsec{sec:experimental}. Then we show our main results in Sections \ref{sec:le_lbeta} and \ref{sec:lmu_ltau}, comparing our sensitivity curves to the leading bounds in the parameter space of the leptophilic models from other experiments.

\subsection{Analysis techniques}\label{sec:experimental}

The Long Baseline Neutrino Facility (LBNF) is expected to produce an intense beam of neutrinos and antineutrinos from a 1.2 MW proton beam colliding against a fixed target \cite{Acciarri:2016crz}. The DUNE ND, where the number of neutrino interactions is largest, is expected to be located at a distance of 574 m from the target. Despite its design not being final yet \cite{Roeck:2018,Manly:2018}, we focus on the possibility of a 75 tonne fiducial mass LAr detector. Regarding the neutrino fluxes, we now concentrate on the option of a beam from 120 GeV protons with $1.1\times10^{21}$ POT per year. The LBNF could also provide higher or lower energy neutrinos depending on the proton energy, target and focusing system used. We explore other possibilities shown in \reftab{tab:rates} and we take the flux files provided in Ref.~\cite{DUNE:flux,DUNE:flux_updated}. We assume that the experiment will run 5 years in neutrino and another 5 years in antineutrino mode. The final exposure, therefore, will vary with beam designs, and is equal to a total of $11\times10^{22}$ POT in the case of 120 GeV protons. To generate neutrino scattering events, we use our own dedicated Monte Carlo (MC), Gaussian smearing the true MC energies and angles as a proxy for the detector effects during reconstruction. We assume an energy resolution of $\sigma/E = 15\%/\sqrt{E}$ $(\sigma/E = 6\%/\sqrt{E})$ for $e/\gamma$ showers (muons) and angular resolutions of $1^\circ$ for all particles \cite{DUNECDRvolII}.

An interesting addition to the design of the DUNE ND would be a magnetized high-pressure gaseous argon tracker placed directly behind the LAr module~\cite{DUNE:GArOption}. The lower thresholds for particle reconstruction and the presence of a magnetic field is expected to improve event reconstruction and reduce backgrounds to neutrino-electron scattering and neutrino trident production. We note that despite the relatively small fiducial mass of such a GAr module, $\lesssim 1$ tonne, it would still provide a sizeable number of these rare leptonic neutrino scattering processes. 

\begin{table}[t]
\begin{center}
\begin{tabular}{|lccccc|}
\hline
		Design & Mode & $\mu^+\mu^-$ trident & $e^+e^-$ trident & $\nu-e$ scattering & POTs/year \\ \hline\hline
		\multirow{ 2}{*}{62.4 GeV $p^+$}       &$\nu$ & $36.5$ & $92.7$ & $7670$ &  $1.83\times10^{21}$\\
		                     &$\overline{\nu}$ & $27.3$ & $73.4$ & $4620$ &  $1.83\times10^{21}$\\\hline
        \multirow{ 2}{*}{80 GeV $p^+$}         &$\nu$ & $42.0$ & $102$ & $8380$ &  $1.4\times10^{21}$\\ 
                             &$\overline{\nu}$ & $33.0$ & $84.3$ & $5320$ &  $1.4\times10^{21}$\\\hline 
        \multirow{ 2}{*}{120 GeV $p^+$}        &$\nu$ & $47.6$ & $110$ & $8930$ &  $1.1\times10^{21}$\\
                             &$\overline{\nu}$ & $40.7$ & $97.6$ & $6450$ &  $1.1\times10^{21}$\\\hline
        \multirow{ 2}{*}{$\nu_\tau$ app optm}  &$\nu$ & $210$ & $321$  & $24900$  & $1.1\times10^{21}$\\
                             &$\overline{\nu}$ & $156$ & $243$  & $14700$  &  $1.1\times10^{21}$\\
        \hline
\end{tabular}
\end{center}
\caption{The SM rates for neutrino trident production and neutrino-electron scattering per year at the 75 tonne DUNE ND after kinematical cuts. \label{tab:rates} }
\end{table}

With the intense flux at DUNE and the large number of POT, the $\nu-e$ scattering measurement will not be statistically limited, with order $10^4$ events in the DUNE ND after a few years. Systematics from the beam and detector are then the limiting factor for the sensitivity to new physics in this measurement. Current work on neutrino flux uncertainties shows that normalization uncertainties can be reduced to the order of $5\%$ \cite{Michael:2008bc,Aliaga:2013uqz,Abramov:2001nr}, with similar projections for DUNE \cite{Acciarri:2016crz}. The electron energy threshold also plays a role in the new physics search. In particular, for new light bosons the enhancement at very low momentum transfer $2 T_e m_e$ has a cut-off at the minimum electron recoil energy (see \refeq{eq:effective_nue_xsec}). This implies that the experiment is no longer sensitive to the $Z^\prime$ mass below $\sqrt{2 T_{\rm th} m_e}$. In our analysis, we assume a realistic overall normalization systematic uncertainty of $5\%$ and $\nu-e$ scattering electron energy thresholds of 600 MeV.

Lowering systematic uncertainties on the flux is challenging given the large hadroproduction and focusing uncertainties at the LBNF beam. Here, improvements on the experimental side in determining the neutrino flux will be extremely valuable (see \eg\ \refref{Bodek:2012uu}). If one is searching for novel leptophilic neutral currents, hadronic processes and inverse muon decay measurements are available, but these are limited either by theoretical uncertainties or by statistics, and might not be applicable in the whole energy region of interest. As to the electron energy, 
assuming a threshold as low as $30$ MeV would be safe for electron detection, but at these low energies backgrounds can be incredibly challenging due to the overwhelming $\pi^0$ backgrounds. Increasing this threshold to $600$ MeV, however, has little impact in our sensitivities and is only $200$ MeV below the threshold used in the most recent MINER$\nu$A analysis \cite{Park:2013dax}, where good reconstruction is important for measuring the flux. For $e^+e^-$ and $\mu^+\mu^-$ tridents, we refrain from increasing the analysis thresholds from a naive $30$ MeV. This is certainly an aggressive assumption but it is necessary if $e^+e^-$ tridents are to be measured, since these events are quite soft \cite{Ballett:2018uuc}. Thresholds for $\mu^+\mu^-$ tridents are much less important since the events are generally more energetic than their dielectron analogue. 

\paragraph{Backgrounds ($\nu_\mu \to \nu_\mu \ell^+\ell^- $)}  We now discuss the individual sources of backgrounds to neutrino trident production. A pair of charged leptons is very rarely produced in neutrino interactions, usually coming from heavy resonance decays \cite{Conrad:1997ne,Astier:2000us,Adams:1999mn,Goncharov:2001qe,PhysRevD.43.2765}. Since our signal is mostly coming from coherent interactions with nuclei, cuts in the hadronic energy deposition in the detector $E_{\rm had}$, often large in heavy meson production processes, can help reduce backgrounds. Coherent and diffractive production of mesons is an exception to this, in particular pion production  \cite{Higuera:2014azj,Mislivec:2017qfz,Acciarri:2014eit,Acciarri:2015ncl}, which is the main background to trident due to particle mis-identification (misID). Muons are known to be easily spoofed by charged pions, making CC $\nu_\mu$ interactions with $\pi^\pm$ in the final state (CC1$\pi$) one of the largest contributions to the backgrounds of $\mu^+\mu^-$ tridents. Similarly, NC $\pi^0$ production stands as the leading background to $e^+e^-$ tridents when the photons are misIDed as two electrons, or if one of the photons pair converts and the other escapes detection. In Ref. \cite{Ballett:2018uuc}, we have shown that the $\mu^+\mu^-$ and $e^+e^-$ pairs produced in trident have small separation angles ($\Delta \theta$), possess small invariant masses ($m^2_{\ell \ell}$) and that both charged leptons are produced with small angles with respect to the neutrino beam ($\theta_{\pm}$). With simplified misID rates, we used the GENIE~\cite{Andreopoulos2009} event generator to show that simple kinematical cuts can reduce backgrounds significantly, achieving a significance of $S_{\mu\mu}/\sqrt{B_{\mu\mu}} \sim 44$ and $S_{ee}/\sqrt{B_{ee}} \sim 17.3$ for the DUNE ND in neutrino mode, where $S$ and $B$ stand for signal and background, respectively. In our current analysis we implement the same kinematical cuts, which are as follows: $m^2_{\mu \mu} < 0.2$ GeV$^2$, $\theta_{\pm} < 15^\circ$ and $\Delta \theta < 20^\circ$ for the $\mu^+\mu^-$ channel, and $m^2_{e e} < 0.1$ GeV$^2$, $\theta_{\pm} < 20^\circ$ and $\Delta \theta < 40^\circ$ for the $e^+e^-$ one. We impose these cuts again in our signal analysis, and point out that the new physics enhancement happens precisely in this favourable kinematical region, (see \refsec{sec:trident_kinematics}). Given the background rejection we find with our naive misID rates, we do not include backgrounds in our results, unless indicated otherwise.

\paragraph{Backgrounds ($\nu-e$)} For neutrino-electron scattering, backgrounds will arise from either the genuine production of an electron or via the misID of particle showers in the detector, both in the absence of observable hadronic energy deposition. The former scenario happens mostly by the CC interactions of the flux suppressed $\nu_e$ states present in the beam. The main contribution will be from CCQE interactions where the struck nucleon is invisible either for being below threshold or due to nuclear re-absorption. The misID of a photon initiated EM shower for an electron one is expected to be rare in LAr, where the first few cm of the showers can be used to separate electrons and photons by their characteristic $dE/dx$. However, the large NC rates for the production of single photons and $\pi^0$ can become a non-negligible background. For instance, coherent NC $\pi^0$ production leaves no observable hadronic signature and may look like a single electron if one of the photons is mis-identified and the other escapes detection. Finally, after misID happens, the signal can still look unique in its kinematical properties. In particular, $E_e \theta^2$ cuts can dramatically reduce backgrounds due to the forwardness of our signal (see e.g. \cite{Park:2013dax,Park:2015eqa}).

\begin{figure}[t]
\centering
\includegraphics[width=\textwidth]{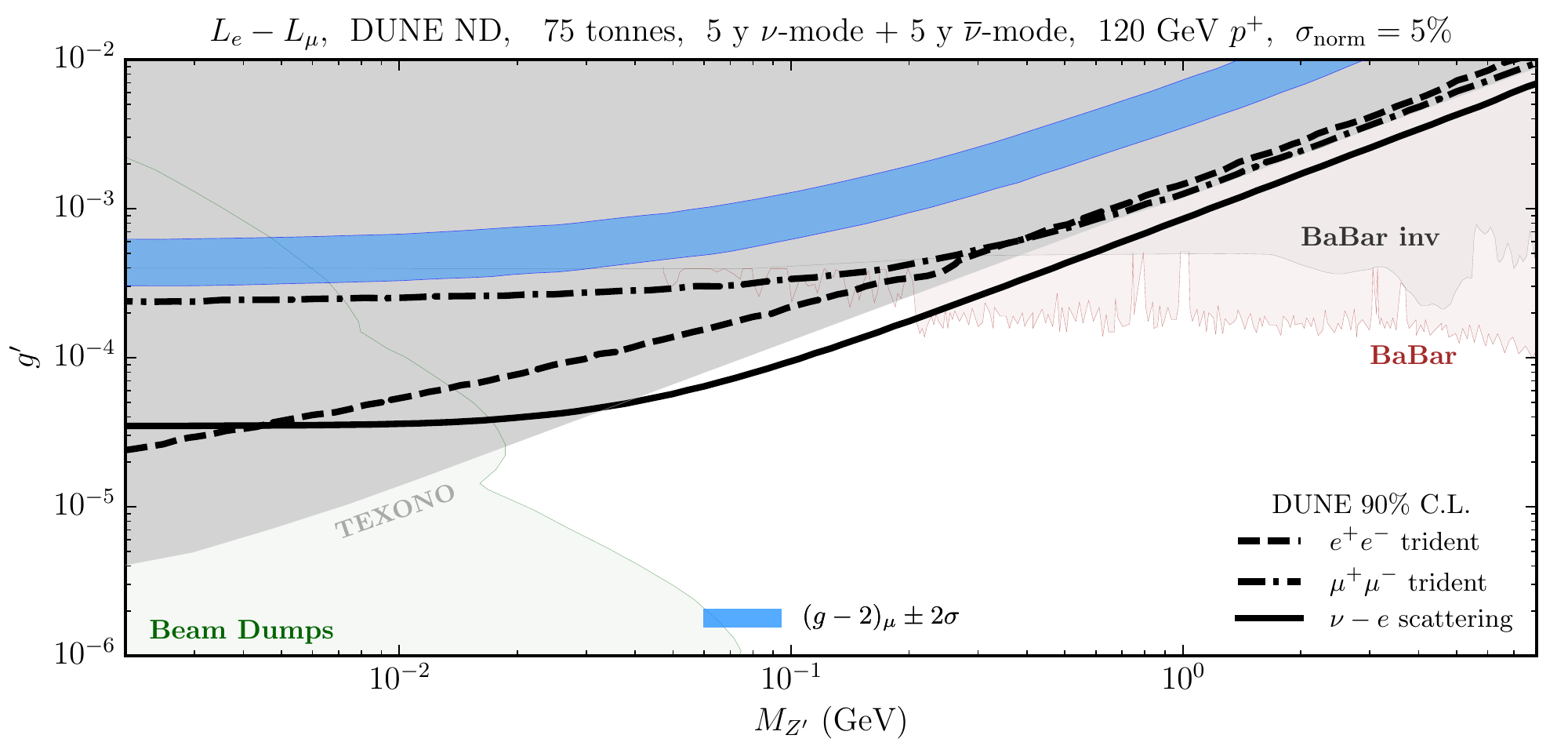}
 \caption{The DUNE near detector neutrino scattering sensitivity to the $L_e - L_\mu$ $Z^\prime$ at 90\% C.L. In solid line we show the $\nu-e$ scattering sensitivity, followed by the dielectron trident in dashed line, and the dimuon trident in dot-dashed line. The coloured regions are excluded by other experiments, where we highlight the neutrino-electron scattering measurements at reactor experiments~\cite{Wong:2006nx,Deniz:2009mu,Chen:2014dsa}, searches at the BaBar $e^+e^-$ collider~\cite{Lees:2014xha, Lees:2017lec} and beam dump experiments~\cite{Bauer:2018onh}.\label{fig:Le_Lmu}}
\end{figure}
\begin{figure}[t]
\centering
    \includegraphics[width=0.49\textwidth]{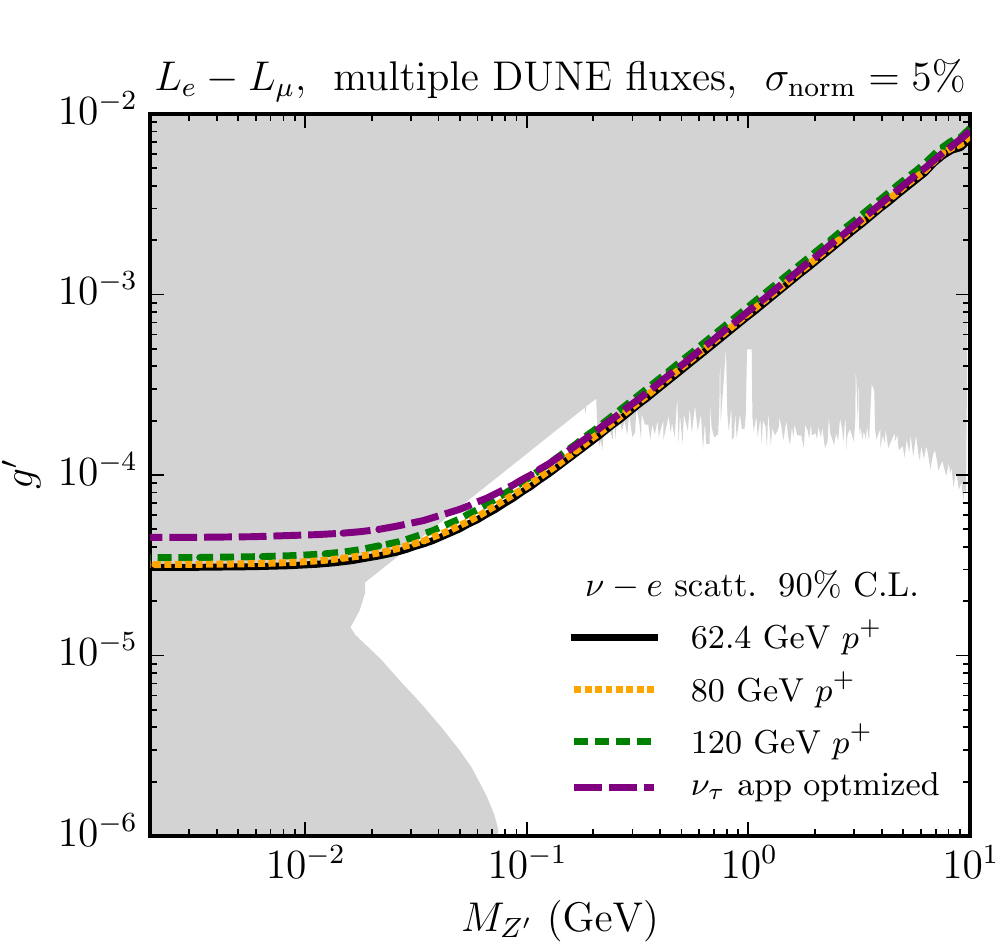}
    \includegraphics[width=0.49\textwidth]{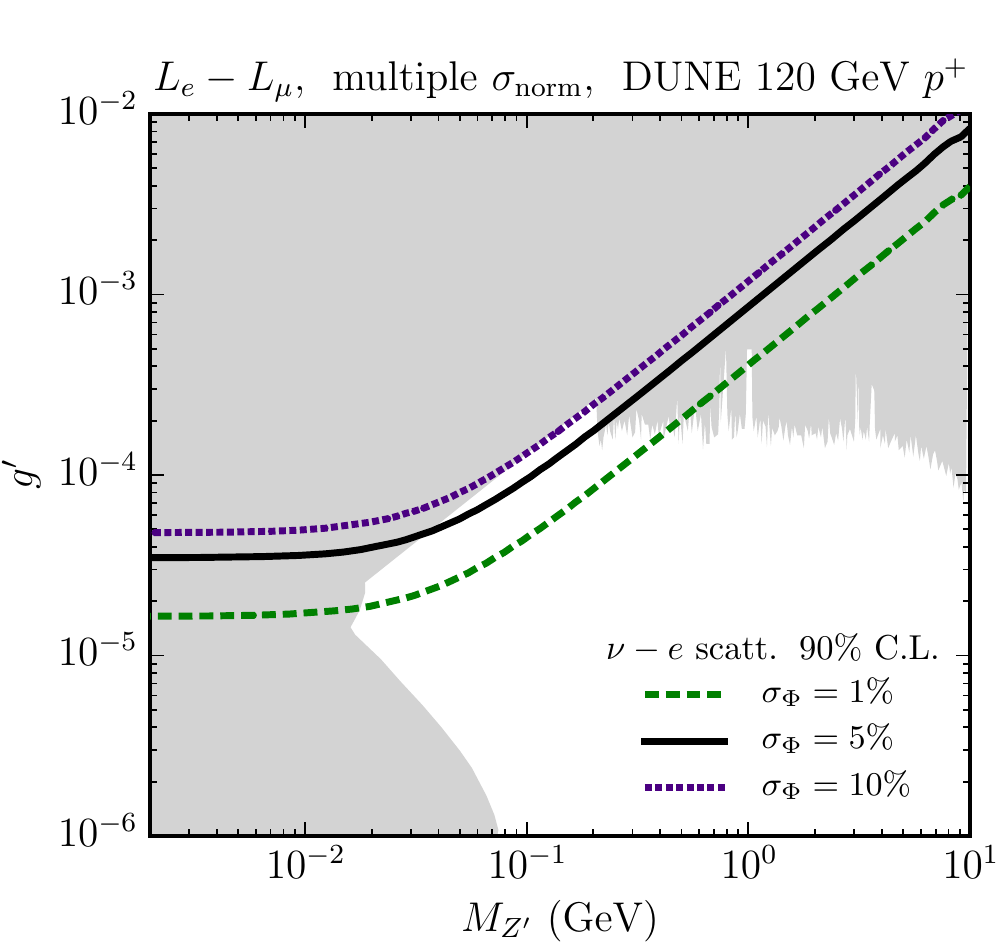}
 \caption{The $\nu-e$ scattering sensitivity to the $L_e - L_\mu$ model at 90\% C.L. On the left panel we show the sensitivity using different choices for the neutrino flux, and on the right we use the neutrino beam from 120 GeV protons and vary the normalization systematic error from an aggressive $1\%$ to a conservative $10\%$ error. \label{fig:Le_Lmu_varied}}
\end{figure}
\paragraph{Statistics.} 
In order to assess the potential of DUNE to discover new physics, we perform a sensitivity analysis using a $\chi^2$ test with a pull method for systematic uncertainties. Our goal is to assess when DUNE would be able to rule out the SM, and so we generate BSM events and fit the SM prediction to it. Our $\chi^2$ function is defined as
\begin{equation}
  \chi^2 = \min_{\alpha} \left[\frac{ (N_{\text{BSM}} - (1+\alpha) N_{\text{SM}}-(\alpha+\beta) N_{\text{BKG}})^2 }{ N_{\text{BSM}}} + \left(\frac{\alpha}{\sigma_{\rm norm}}\right)^2 + \left(\frac{\beta}{\sigma_{\rm BKG}}\right)^2 \right],
\end{equation}
where the number of events for the BSM case is given by $N_{\rm BSM}$, the SM number of events is $N_{\rm SM}$ and the number of background events is $N_{\rm BKG}$. The nuisance parameters $\alpha$ and $\beta$, with their uncertainties $\sigma_{\rm norm}$ and  $\sigma_{\rm BKG}$, take into account normalization uncertainties from the flux and detector, and uncertainties on the background prediction, respectively. For the DUNE ND, we assume $\sigma_{\rm norm} = 5\%$ and $\sigma_{\rm BKG} = 10\%$. These systematics will likely be dominated by flux normalization uncertainties, and can only be measured with interactions that do not depend on the leptophilic BSM physics.

\subsection{\boldmath${L_e - L_\mu}$ \label{sec:le_lbeta}}

New vector bosons with couplings to the first and second generation leptons can be probed very effectively in neutrino experiments by meausuring the $\nu-e$ scattering rate. This has been recognized in the literature \cite{Bilmis:2015lja,Lindner:2018kjo,Bauer:2018onh}, where bounds from various experiments, including CHARM-II \cite{Vilain:1994qy}, TEXONO \cite{Wong:2006nx,Deniz:2009mu,Chen:2014dsa} and Borexino \cite{Bellini:2011rx} have been derived on these bosons. Curiously, the bound calculated from the CHARM-II data has been pointed out by \refref{Bauer:2018onh} to be too optimistic. The uncertainty on the neutrino flux is a real hindrance for these measurements which has not been taken into account when these bounds were computed. This is particularly important for measurements with large statistics, and for this reason we do not show the CHARM-II bound here. The measurement of $\overline{\nu}_e - e$ scattering at TEXONO, on the other hand, is statistically limited, and the bound it places on this class of models can safely ignore the flux systematics. This turns out to provide the strongest limit in a large region of the ${L_e - L_\mu}$ parameter space. Trident bounds can be obtained for this model, but due to their lower statistics and more involved kinematics, are subdominant.

We show our results for the DUNE ND in \reffig{fig:Le_Lmu}. Our results are for the combined $\nu+\bar\nu$ modes and do not include backgrounds. The opposite charges between the first and second families implies constructive interference between the SM and BSM contributions for neutrino scattering, contrary to what happens in a $B-L$ model, for instance. Therefore, the strongest bounds on this model can be obtained at DUNE in neutrino mode. It is clear, however, that the degree with which DUNE can probe unexplored parameter space is a question on how much the uncertainties on the flux can be lowered. To illustrate this effect, we vary the normalization systematics on the right panel of Fig.~\ref{fig:Le_Lmu_varied}, going from a conservative $10\%$ to an aggressive $1\%$ uncertainty. The effect of changing the thresholds is very small, being most important in a region already probed by other experiments. Different beam designs could also have an impact on the sensitivity, although this is quite small, as shown on the left panel of Fig.~\ref{fig:Le_Lmu_varied}.

Since we show the bounds obtained from the neutrino and antineutrino runs combined, it is not possible to see effects of destructive interference. If only channels with destructive interference are available, however, it would have been possible to allow for cancellations between the total interference and the square of the BSM contributions in certain regions of parameter space at the level of the total rate. The region where this cancellation happens depends strongly on the neutrino energies involved and on the integrated phase space of the recoiled electron. In that case, one expects that the sensitivity to the lowest new physics couplings comes, in fact, from the search for a deficit of $\nu-e$ scattering events, as opposed to the constructive interference case where an excess of events is always produced. We note that this has no significant impact in the sensitivity of a leptophilic $Z^\prime$, but might provide crucial information about the nature of the $Z^\prime$ charges in case of detection. 

The trident bounds we obtain are not competitive for this model despite the fact that the trident cross sections receive similar enhancements to that of $\nu-e$ scattering. This is due to two reasons: the low number of events and the non-trivial kinematics of trident processes. Since the neutrino is essentially scattering off virtual charged leptons produced in the Coulomb field of the nucleus, it has to typically transfer more energy to the system than it would in a scattering off real particles in order to produce visible signatures. This remark also helps us explain the behaviour of the sensitivity curves at the lowest masses. Whilst $\nu-e$ scattering cross sections become insensitive to the boson mass at $\sqrt{2 m_e T_{\rm th}}$, the trident cross sections do not. This behaviour is most dramatic in the $e^+e^-$ tridents, but is also present in the $\mu^+\mu^-$ one. This is a consequence of the 4-body phase space kinematics, where now the momentum transfer through the $Z^\prime$ propagator is no longer trivially related to the final state particle energies, as in $2\to2$ processes. It should be noted, however, that both the dimuon and the dielectron trident rates become nearly independent of $M_{Z^\prime}$ below the muon and the electron mass, respectively, where only a logarithmic dependence is expected~\cite{Altmannshofer2014}.

DUNE can also proble this class of models in a different way. In the context of long range forces in neutrino oscillation experiments and with the same choice of charges, Ref. \cite{Wise:2018rnb} places competitive bounds in this model with Super-Kamiokande data and makes projections for DUNE. The matter potential created by the local matter density modifies the dispersion relation of the neutrinos with lepton non-universal charges, leading to very competitive bounds in our region of interest. Other experimental searches have been conducted at electron beam dumps. This technique consists of producing the $Z^\prime$ boson at the target via radiative processes such as $e + A \to e + A + Z^\prime$, and look for the visible decays of the boson in the detector. In this model, the decay products are mostly $e^+ e^-$ states and the bounds are only applicable at appreciably small values of $g^\prime$ and $M_{Z^\prime}$, where the lifetime of the $Z^\prime$ is sufficiently large. Probing the large mass region, on the other hand, requires high-energy experiments. In that regime, the strongest bounds come from searches at the $e^+ e^-$ collider BaBar. These come about in two ways: looking for the visible decay products of a $Z^\prime$ produced radiatively or in heavy meson decays \cite{Lees:2014xha}, or exploring the BR into invisible final states \cite{Lees:2017lec}.
\begin{figure}[t]
\centering
  \includegraphics[width=\textwidth]{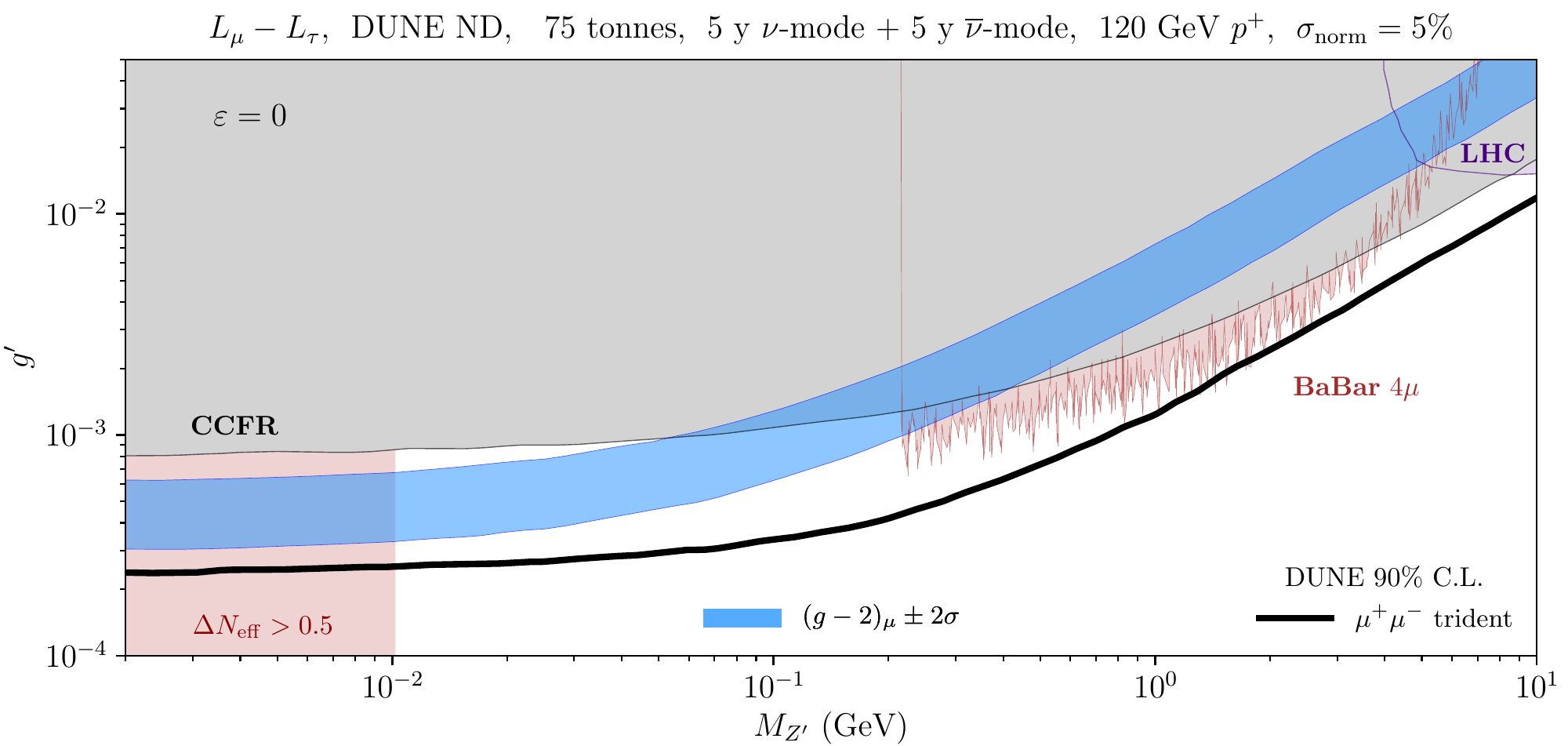}\\
 \includegraphics[width=\textwidth]{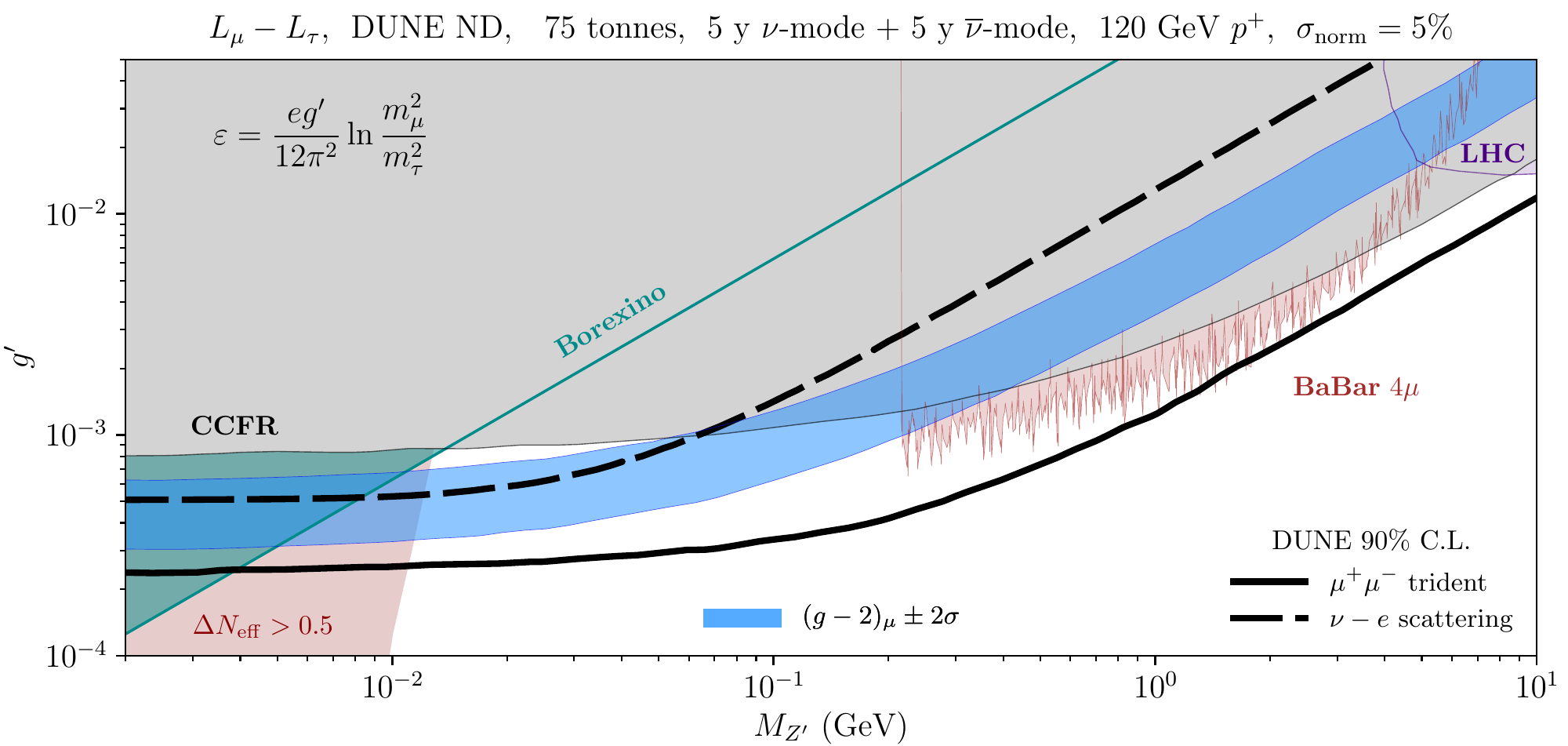}
 \caption{The DUNE near detector neutrino scattering sensitivity for $L_\mu - L_\tau$ at 90\% C.L. The upper panel shows the case with no kinetic mixing, and the lower panel the case with the loop-induced mixing. Bounds from neutrino-electron scattering apply only to the latter. We also show bounds from BaBar~\cite{TheBABAR:2016rlg}, LHC~\cite{Aad:2014wra}, Borexino~\cite{Kaneta:2016uyt} and from the neutrino trident production measurement at CCFR~\cite{Mishra:1991bv,Altmannshofer2014}. Recent cosmological bounds for the two kinetic mixing cases derived in Ref.~\cite{Escudero:2019gzq} are also shown. \label{fig:Lmu_Ltau}}
\end{figure}
\begin{figure}[t]
\centering
  \includegraphics[width=0.49\textwidth]{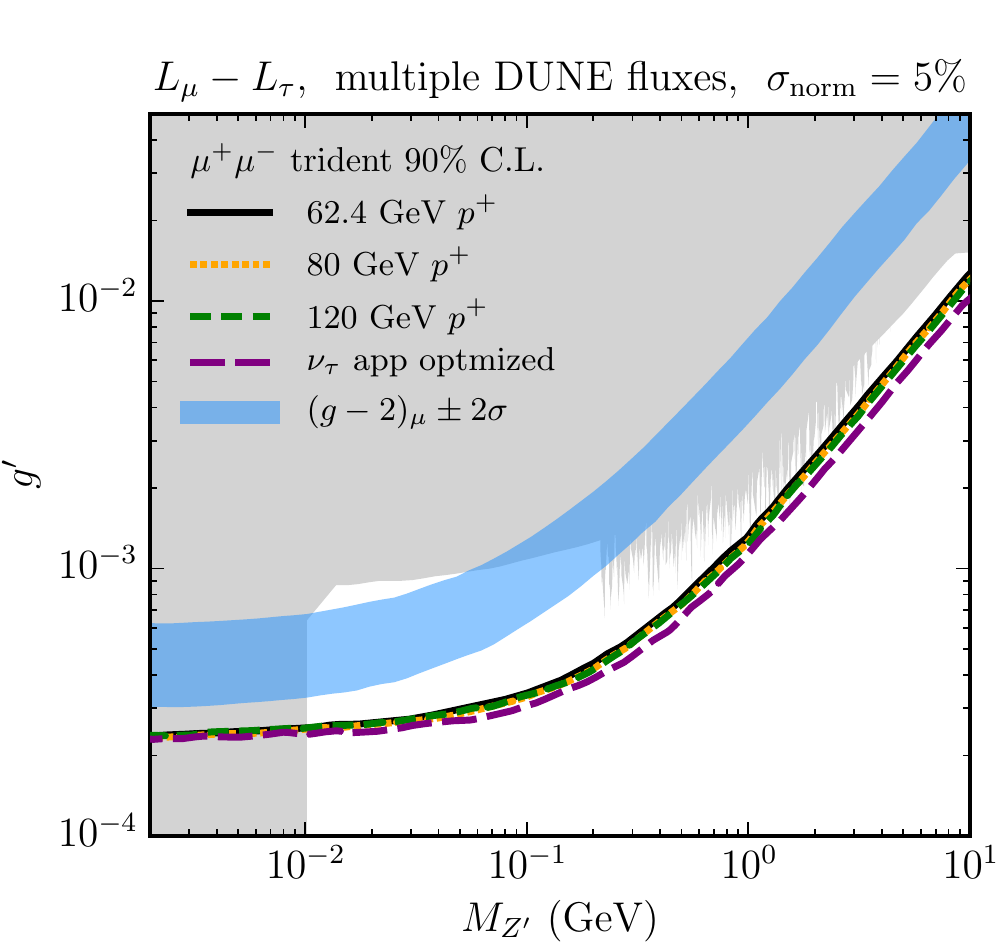}\textbf{}
 \includegraphics[width=0.49\textwidth]{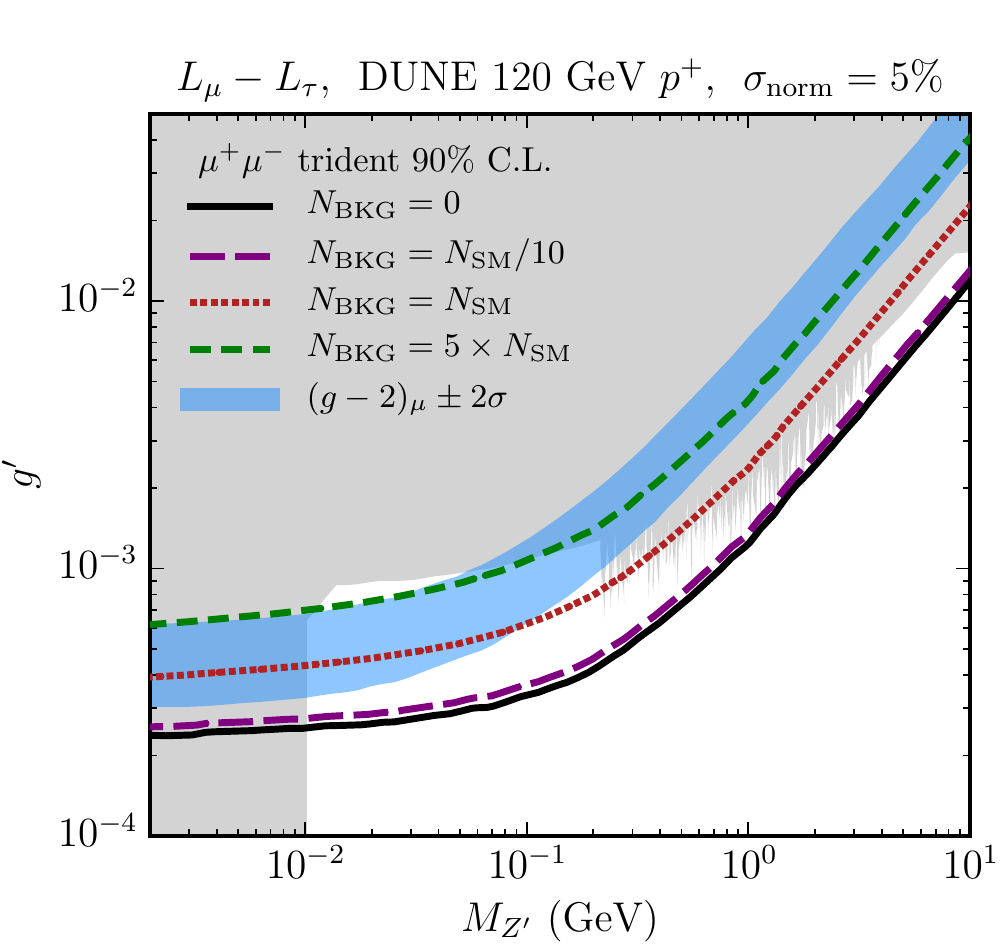}
 \caption{The dimuon neutrino trident sensitivity to the $L_\mu - L_\tau$ model with no kinetic mixing at 90\% C.L. On the left panel we show the sensitivity using different choices for the neutrino flux, and on the right we use the neutrino beam from 120 GeV protons and scale the background with respect to the total number of SM trident events after cuts.\label{fig:Lmu_Ltau_varied}}
\end{figure}
%

\subsection{\boldmath$L_\mu - L_\tau$ \label{sec:lmu_ltau}}

In this section we evaluate the DUNE ND sensitivity to the presence of a light
vector $Z^\prime$ charged under $L_{\mu} - L_{\tau}$. Beyond being anomaly free, this choice of charges allows for positive contributions to the anomalous magnetic moment of the muon, $a_\mu = (g-2)_{\mu}$, as discussed in Refs. \cite{Baek:2001kca,Pospelov:2008zw,Kamada:2015era,Araki:2015mya}. This quantity is well known for a $\sim 3.7\sigma$ discrepancy between the experimental measurement \cite{Bennett:2006fi} and the theory predictions \cite{Blum:2018mom,Keshavarzi:2018mgv}. If future efforts to measure it \cite{Grange:2015fou} confirm this disagreement and if theoretical uncertainties are better controlled in the next few years, then constraining new physics scenarios that could contribute to $a_\mu$ is of utmost importance. 

This model can significantly impact neutrino trident production of a muon pair. In fact, the leading bound in this parameter space for masses $M_{Z^\prime} \lesssim 200$ MeV comes from the CCFR measurement of the same neutrino trident channel \cite{Mishra:1991bv}. CCFR observed $37.0\pm 12.4$ events, extracting a measurement of the trident cross section of $\sigma_{\rm CCFR}/\sigma_{\rm SM} = 0.82 \pm 0.28$. Curiously, the measurement by CHARM-II \cite{Geiregat:1990gz} provides weaker constraints on this model despite seeing a larger number of trident events, namely $55 \pm 16$ events in total, most likely due to the $1\sigma$ upward fluctuation in the measurement: $\sigma_{\rm CHARM-II}/\sigma_{\rm SM} = 1.58 \pm 0.57$. Other important bounds from $\nu-e$ scattering have also been obtained using the kinetic mixing parameter generated at one-loop. The strongest of which uses data from Borexino \cite{Kaneta:2016uyt}, and are only relevant for the lowest mass region $M_{Z^\prime} \lesssim 20$ MeV.

At DUNE, both of these measurements are possible, allowing to constrain this model in different ways. We show our results in Fig.~\ref{fig:Lmu_Ltau}, without including backgrounds. In this scenario, DUNE would be able to cover all the $2 \sigma$ region of the $(g-2)_\mu$ measurement only with the measurement of $\mu^+ \mu^-$ tridents. For the low mass region, measuring the $\nu-e$ scattering rate can provide a complementary probe of this region, depending most strongly on the systematic uncertainties DUNE can achieve. We note that analysis thresholds used for $\nu-e$ scattering have little impact on the sensitivity in the region of interest. Our conclusion that DUNE can cover all of the $(g-2)_\mu$ region holds provided backgrounds are kept below the SM signal rate. This can be seen when we include backgrounds with different assumption on the right panel of Fig.~\ref{fig:Lmu_Ltau_varied}. Finally, different assumption for the beam design have little impact on the sensitivity, as show on the left panel of Fig.~\ref{fig:Lmu_Ltau_varied}.

Apart from neutrino scattering, dedicated searches for resonances decaying into $\mu^+ \mu^-$ in four muon final states have been performed at the BaBar \cite{TheBABAR:2016rlg}, $e^+ e^- \to \mu^+\mu^- Z^\prime (\to \mu^+ \mu^-)$. At the LHC, the searches for $Z \to 4 \mu$ \cite{Aad:2014wra} performed by the ATLAS collaboration can be recast into a bound in the $Z^\prime$. Here, we show the bounds derived in \cite{Altmannshofer2014}. Big Bang Nucleosynthesis bounds were studied in \cite{Kamada:2015era}, and shown to constrain the mass of the boson to be $M_{Z^\prime} \gtrsim 5$ MeV. Recently, additional constraints from Cosmology were derived given that the presence of very light $Z^\prime$ bosons changes the evolution of the early Universe~\cite{Escudero:2019gzq}. In particular, the decays and inverse decays induced by the new leptophilic interactions can modify the neutrino relativistic degrees of freedom, requiring $M_{Z^\prime}\gtrsim 10$ MeV in order for $\Delta N_{\rm eff} < 0.5$ for the case with no kinetic mixing. The authors of Ref.~\cite{Escudero:2019gzq} also found that an additional $Z^\prime$ boson can alleviate the tension in the different measurements of the Hubble parameter. Let us stress here that all these bounds will be complementary to possible future constraints that can be obtained by the DUNE program, as shown in Fig. \ref{fig:Lmu_Ltau}.

\section{\label{sec:conclusion}Conclusions}

Although next generation oscillation experiments are designed for making precision measurements of the neutrino mixing parameters, the unprecedented fluxes and large detectors will allow for many non-minimal new physics searches. In this work, we have considered the physics potential of the near detector of DUNE for constraining the existence of additional anomaly-free $U(1)$ gauge group which only couple to leptons --- a form of purely leptophilic neutral current. Specifically, we have considered the anomaly free scenarios with charges associated to the lepton number difference $L_\alpha-L_\beta$.  Focusing on the two most promising neutrino scattering processes, $\nu- e$ scattering and ($\nu\ell\ell$) trident scattering, we have computed expected sensitivity curves for the DUNE ND for a variety of charge assignments. 

In performing our sensitivity studies to the coupling and mass of the $Z^\prime$ boson, we have remained as faithful as possible to the real experimental conditions of a LAr detector. Our main results rely on the realistic assumptions of flux uncertainties of $5\%$ and feasible exposures. To avoid large backgrounds, we have also implemented kinematical cuts on the neutrino trident sample, and an analysis threshold of $600$ MeV for neutrino-electron scattering. The parameter space which can be probed by neutrino-electron scattering in the $L_e-L_\mu$ scenario is at least two times better than $e^+e^-$ and almost twenty times better than the $\mu^+\mu^-$ trident channels, specially for the lower mass region. In this case, the DUNE ND would improve only slightly on previous neutrino-electron scattering bounds, especially at around $M_{Z^\prime}\sim100$ MeV. We do not expect $e^+e^-$ trident measurements at DUNE to improve our coverage of the $L_e - L_\mu$ $Z^\prime$ parameter space, but note this process has a distinct dependence on $M_{Z^\prime}$ if compared to $\nu-e$ scattering. 

If the light vector $Z^\prime$ is charged under $L_\mu-L_\tau$, we have found that the dimuon trident measurement could provide the leading bound in this parameter space. This is particularly interesting as these models can also explain the discrepancy between the measurement of the anomalous magnetic moment of muon and its SM prediction. We expect that DUNE will be able to fully explore the $(g-2)_\mu$ motivated parameter space provided backgrounds are kept under control. The robustness of our results is tested against different choices of neutrino fluxes, where we find that despite the larger rates at higher neutrino energies and the larger BSM enhancement at lower energies, the sensitivities are very similar. 

Improvements to the experimental sensitivities we have displayed in Figs.~\ref{fig:Le_Lmu} and \ref{fig:Lmu_Ltau} can be achieved by reducing uncertainties on the neutrino flux and detection. From the experimental side, novel detection techniques suitable to rare neutrino events are currently under discussion, such as the magnetized HPgTPC~\cite{DUNE:GArOption} and the Straw Tube Tracker~\cite{PettiTalk,Duyang:2018lpe}. Together with improved analysis techniques, these will help to improve upon our projections for the sensitivity of DUNE to new physics that might be hiding at light masses and small couplings.

\noindent\textbf{Note added}: During the completion of this work, Ref.~\cite{Altmannshofer:2019zhy} appeared with similar results for the sensitivity of DUNE to $L_\mu-L_\tau$ bosons using $\mu^+\mu^-$ tridents.


\acknowledgments

The authors would like to thank TseChun Wang for his contribution during the early stages of the work. We are also grateful for stimulating discussions with Martin Bauer. This work was partially supported by Funda\c{c}\~{a}o de Amparo \`{a} Pesquisa do Estado de S\~{a}o Paulo (FAPESP) and Conselho Nacional de Ci\^{e}ncia e Tecnologia (CNPq). This project has also received support from the European Union's Horizon 2020 research and innovation programme under the Marie Sk\l odowska-Curie grant agreement No. 690575 (RISE InvisiblesPlus) and No. 674896 (ITN Elusives). SP and PB are supported by the European Research Council under ERC Grant NuMass (FP7-IDEAS-ERC ERC-CG 617143). SP
acknowledges partial support from the Wolfson Foundation and the Royal Society. Fermilab is operated by the Fermi Research Alliance, LLC under contract No. DE-AC02-07CH11359 with the United States Department of Energy.

\appendix

\section{Trident phase space \label{app:phase_space}}

In this appendix we derive a phase space parametrization for neutrino trident production in terms of the momentum transfer $K^2 =  2 p_1 \vdot p_2$. This is important if one wants to change variables to smooth out the integrand at low $M_{Z^\prime}$ masses. We follow the calculation in \cite{Czyz1964} and \cite{Ballett:2018uuc}, and proceed to define $K^2$ as one of the integration variables. The relevant lorentz invariant phase space for the $2\to3$ leptonic part of the cross section is given by
\begin{equation}
\int \dd^3 \Pi_{\mathrm{LIPS}} = \int \frac{\dd \vec{p_2} }{(2\pi)^32 E_2} \frac{ \dd \vec{p_3} } {(2\pi)^3 2 E_3} \frac{\dd \vec{p_4}}{(2\pi)^32 E_4} \, (2\pi)^4\delta^{(4)} (p_1 + q - p_2 - p_3 - p_4).
\end{equation}
Following \cite{Czyz1964} we start by working in the frame $\vec{p_1} + \vec{q} - \vec{p_3} = 0$, putting $\vec{p_1}$ along the $\hat{z}$ direction instead. The delta function can be integrated with the $\vec{p_4}$ and $|\vec{p_2}|$ integrals, such that 
\begin{equation}
 \int \frac{\dd \vec{p_2} }{2 E_2} \frac{\dd \vec{p_4}}{2 E_4} \, \delta^{(4)} (p_1 + q - p_2 - p_3 - p_4) =  \int \frac{ |\vec{p_2}|}{4 W_c} \, \frac{1}{E_1 E_2} \dd K^2 \, \dd \phi_2,
\end{equation}
where we defined
\begin{align}
|\vec{p_2}| = (W_c^2 - m_1^2)/2W_c, \quad W_c = q^0 + E_1 - E_3, \quad K^2 = 2 E_1 E_2 (1 - \cos{\theta_2}).
\end{align}
Since we conserve energy and momentum in this frame, we can take $-1 \leq \cos{\theta_2} \leq 1$ and $0 \leq \phi_2 \leq 2 \pi$. The remaining $\vec{p_3}$ integral can be performed with the variables defined in \cite{Czyz1964} to yield
\begin{equation}
 \int \frac{\dd \vec{p_3} }{2 E_3}  =  \int \frac{2 \pi}{\hat{s}} \, \dd x_5 \, \dd x_3,
\end{equation}
where a trivial azimuthal angle was integrated over. Their limits are more easily found in the frame $\vec{p_1} + \vec{q} = 0$, with $\vec{q}$ along the $\hat{z}$ direction. Finally, our main result is given by
\begin{equation}
\int \dd^3 \Pi_{\mathrm{LIPS}} = \frac{1}{(2\pi)^4} \int \frac{|\vec{p_2}|}{4 W_c} \, \frac{1}{\hat{s}} \, \frac{1}{E_1 E_2} \, \dd x_5 \, \dd x_3 \, \dd K^2 \, \dd \phi_2 .
\end{equation}
There remains two non-trivial integrations to be performed to obtain the full four body phase space cross section, namely the ones over $q^2$ and $\hat{s}$. The substitutions suggested in \cite{Lovseth1971} for these two invariants are still convenient, and we make use of these in our numerical integrations.

\section{Weak form factor}
\label{app:formfactors}

Here we show our weak hadronic current used in the dark-bremsstrahlung calculation. Similarly to the electronmagnetic case, we write the weak hadronic current for a spin-0 nucleus with $Z$ protons and $N$ neutrons as
\begin{equation}
 {\rm H}_{\rm W}^\mu = \bra{\mathcal{H}(k_3)} J^\mu_{\rm W} (Q^2) \ket{\mathcal{H}(k_b)} = Q_{\rm W} (k_b+k_3)^\mu F_{\rm W}(Q^2),
\end{equation}
where $Q_{\rm W} = (1-4 s_{\rm w}^2)Z - N$ and $F_{\rm W}(Q^2)$ stands for the weak form factor of the nucleus. We implement the Helm form factor as in \cite{Duda:2006uk}, defined as  
\begin{equation}
|F(Q^2)|^2 = \left( \frac{3 j_1(QR)}{QR}\right)^2 e^{-Q^2 s^2},
\end{equation}
where $j_1 (x)$ stands for the spherical Bessel function of the first kind, $s = 0.9$ fm and $R = 3.9$ fm for $^{40}$Ar.

\bibliographystyle{JHEP}
\bibliography{main}{}

\end{document}

%% file: commands.tex
\newcommand{\refeq}[1]{Eq.~(\ref{#1})}

\newcommand{\reffig}[1]{Fig.~\ref{#1}}

\newcommand{\refsec}[1]{Section~\ref{#1}}
\newcommand{\refapp}[1]{Appendix~\ref{#1}}
\newcommand{\reftab}[1]{Table~\ref{#1}}
\newcommand{\refref}[1]{Ref.~\cite{#1}}
\newcommand{\refrefs}[2]{Refs.~\cite{#1}~and~\cite{#2}}

\def\lagrangian{lagrangian}
\def\eg{\emph{e.g.}}

\definecolor{light-gray}{gray}{0.65}

\newcounter{CommentCount}
\definecolor{PB}{rgb}{0.9,0,0}
\definecolor{MH}{rgb}{0.0,0.0,0.5}
\definecolor{YP}{RGB}{68,50,127}